\documentclass[
    twocolumn,
	aps,
	prd,
	superscriptaddress,
	preprintnumbers,
	secnumarabic,
	nofootinbib]{revtex4-1}
  
\pdfoutput=1

\usepackage{graphicx}
\usepackage{enumitem}
\usepackage{latexsym}
\usepackage{amsfonts}
\usepackage{amssymb}
\usepackage{color}
\usepackage{xcolor}
\usepackage{amsmath}
\usepackage{slashed}
\usepackage{dcolumn}
\usepackage{verbatim}
\usepackage{float}
\usepackage{xspace}

\newcommand{\half}{\frac{1}{2}}
\newcommand{\gsim}{\gtrsim}
\newcommand{\lsim}{\lesssim}
\newcommand{\ra}{\rightarrow}

\newcommand{\ct}{c_\theta}

\def\sigmaveeave{\langle \sigma v\rangle}
\def\sigmaveeann{\langle \sigma v\rangle_{\rm ann}}

\def\Lc{\mathcal{L}}
\def\Oc{\mathcal{O}}
\def\Mc{\mathcal{M}}

\def\met{\slashed{E}_T}

\newcommand{\beq}{\begin{equation}}
\newcommand{\eeq}{\end{equation}}
\newcommand{\bea}{\begin{eqnarray}}
\newcommand{\eea}{\end{eqnarray}}
\newcommand{\nn}{\nonumber}

\def\qmed{\tilde{Q}}
\def\lmed{\tilde{L}}
\def\lambdaq{\lambda_{\tilde{Q}}}
\def\lambdal{\lambda_{\tilde{L}}}
\def\ace{A_{\rm CE}}
\def\afb{A_{\rm FB}}

\def\mll{m_{\ell \ell}}
\def\mee{m_{ee}}

\def\mchi{m_\chi}
\def\mdm{m_\chi}
\def\mphi{m_\phi}
\def\mmed{m_\phi}

\def\delm{\delta m}

\def\ctCS{\cos\theta_{\rm CS}}
\def\ct{\cos\theta}
\def\fRR{\tt pD^u_{\rm RR}}
\def\fRL{\tt pD^u_{\rm RL}}
\def\sRR{\tt pCS^u_{\rm RR}}
\def\sRL{\tt pCS^u_{\rm RL}}
\def\fRRd{\tt pD^d_{\rm RR}}
\def\fRLd{\tt pD^d_{\rm RL}}
\def\sRRd{\tt pCS^d_{\rm RR}}
\def\sRLd{\tt pCS^d_{\rm RL}}

\newcommand{\acro}[1]{\textsc{\MakeLowercase{#1}}}   
\newcommand{\DM}{\acro{DM}\xspace}
\newcommand{\SM}{\acro{SM}\xspace}
\newcommand{\LHC}{\acro{LHC}\xspace}
\newcommand{\ATLAS}{\acro{ATLAS}\xspace}
\newcommand{\CMS}{\acro{CMS}\xspace}
\newcommand{\QCD}{\acro{QCD}\xspace}
\newcommand{\NLO}{\acro{NLO}\xspace}
\newcommand{\MET}{\acro{MET}\xspace}
\newcommand{\CL}{\acro{C.L.}\xspace}
\newcommand{\LSP}{\acro{LSP}\xspace}

\definecolor{orange}{rgb}{1,0.5,0}
\definecolor{purple}{rgb}{1,0,1}
\definecolor{brown}{rgb}{.7,.2,.2}
\definecolor{violet}{rgb}{.6,.3,.8}

\allowdisplaybreaks

\begin{document}

\title{Characterizing dark matter at the LHC in Drell-Yan events}

\author{Rodolfo M. Capdevilla}
\author{Antonio Delgado}
\author{Adam Martin}
\author{Nirmal Raj}

\affiliation{Department of Physics, University of Notre Dame, 225 Nieuwland Hall, Notre Dame, Indiana 46556, USA}

\begin{abstract}

Spectral features in \LHC dileptonic events may signal radiative corrections coming from new degrees of freedom, notably dark matter and mediators.
Using simplified models, we show how these features can reveal the fundamental properties of the dark sector, such as self-conjugation, spin and mass of dark matter, and the quantum numbers of the mediator.
Distributions of both the invariant mass $\mll$ and the Collins-Soper scattering angle $\ctCS$ are studied to pinpoint these properties.
We derive constraints on the models from \LHC measurements of $\mll$ and $\ctCS$, which are competitive with direct detection and jets + $\met$ searches.
We find that in certain scenarios the $\ctCS$ spectrum provides the strongest bounds, underlying the importance of scattering angle measurements for non-resonant new physics. 
\end{abstract}
                 
\maketitle

\section{Introduction}

Astrophysical evidence for dark matter (\DM) abounds, but its fundamental properties remain elusive. 
Key puzzles that remain unsolved are:

\begin{itemize}

\item Is \DM its own antiparticle?

\item Does it carry spin? 

\item What is its mass?

\item How does it couple to the Standard Model, if at all? 

\end{itemize} 
These properties result in qualitatively diverse signals in direct detection searches (such as whether scattering is spin-independent, spin-dependent and/or momentum-dependent) and indirect detection (such as whether annihilation is $s$-wave or $p$-wave). 
Thus, tests may devised by which these properties may be marked out \cite{0803.4477,0808.3384,1110.4281,1305.1611,1506.04454,1610.06581,1706.07819}.
In this paper, we ask if the same can be done at a collider, compelled by the fact that Run 2 of the Large Hadron Collider (\LHC) is well underway.
We find a surprising lack in the literature of \LHC-related work addressing the questions of self-conjugation, spin and coupling structure, perhaps because the primary focus of most collider searches is to extract the mass of \DM and possibly that of a mediator that couples the \DM particle to the \SM.
Two exceptions are Ref.~\cite{1610.07545}, where spin-1/2 and spin-1 \DM were distinguished using distributions of $\met$, jet rapidity, and \DM invariant mass, and Ref.~\cite{1706.04512}, where \DM properties were distinguished by decomposing the missing energy spectrum into basis functions. These studies make use of $\met$, the most striking feature of \DM directly produced on-shell at the \LHC.

In this work, we will focus on collider signals that can potentially address these questions, but take an approach that is {\em not} $\met$-based. 
Instead, we ask if event distributions of {\em fully visible} final states can hold the key. 
A dark sector can leave its imprint in visible spectra if it induces loop processes interfering with Standard Model (\SM) amplitudes; in particular, threshold effects may generate distinct signal features.
As shown in Ref.~\cite{1411.6743}, such non-resonant signals are best discernible in $\ell^+ \ell^-$ production at the \LHC: the backgrounds are simple and intelligible, the rates are high, and the events are precisely reconstructed\footnote{These process features have also been exploited for probing $R$-parity violation \cite{Choudhury:2002av}, running of electroweak couplings \cite{Rainwater:2007qa,Alves:2014cda,Gross:2016ioi}, electroweak precision observables \cite{1609.08157}, and leptoquarks \cite{Hewett:1997ce,Wise:2014oea,Bessaa:2014jya,1610.03795}, not to mention the ubiquitous literature on resonant $Z'$ bosons.}.  
Indeed, the channel is so clean that in some regions it turns out to be more sensitive to the dark sector parameters than conventional jets + $\met$ and direct detection searches.
Note that the collider signals are agnostic to the \DM abundance, and can be relevant for models that populate a fraction $< 1$ of \DM via thermal freezeout.


In the current paper we extend the program of Ref.~\cite{1411.6743} to study how well dileptonic information can shed light on the quantum properties of the dark sector. 
As in~\cite{1411.6743}  we will construct ``simplified models" where  \DM couples to both quarks and leptons, for which we introduce mediators charged appropriately.
The simplified frameworks are 
renormalizable effective theories characterized by a minimal set of inputs, usually no more than the \SM-\DM coupling, the masses of \DM and the mediator, and specifications of \DM spin and the mediator's quantum numbers \cite{1307.8120,1308.0592,1308.0612,1308.2679,1402.2285,1403.4634,1409.2893,1410.6497,1411.0535,1506.03116,1510.03434,1608.05345}. 
While Ref.~\cite{1411.6743} focused on Dirac \DM that only coupled to right handed \SM fermions, we will survey and compare several scenarios: \DM that is self-conjugate and not, \DM with spin 0 and spin 1/2, \DM that couples to right-handed fermions and left-handed. 
We make full use of the information available in dileptonic events at the \LHC, meaning we study spectra of the invariant mass ($\mll$) {\em and} scattering angle.
To our knowledge, this is the first paper availing \LHC measurements of dilepton angular spectra to probe \DM and its mediators.
In fact, we find that angular spectra may provide the strongest constraints. 
This statement is not entirely surprising; non-resonant new physics must produce comparable effects on both the invariant mass and angular distributions, and so it is reasonable to expect the latter to sometimes have more sensitivity \cite{1610.03795}.

Our paper is laid out as follows.
In Sec.~\ref{sec:models}, we introduce the \DM models we will use for our study and state our simplifying assumptions.
In Sec.~\ref{sec:signals}, we show the various signals produced by our set-up and discuss how their features may help distinguish between our models. Next, in Sec.~\ref{sec:pheno}, we discuss all the relevant constraints on our models, comparing the dileptonic probes with jets + $\met$, relic density and direct detection constraints.
We also show the future prospects of our set-up at the \LHC at high luminosity.
In Sec.~\ref{sec:disc} we summarize our findings and conclude.

\begin{table*}[t]
\begin{center}
\begin{tabular}{l l l l l}
Model~~~~ & $\chi$ spin~~~~ & $\qmed, \lmed$ spin~~~~ & $\qmed$ under $G_{\rm SM}$~~~~ & $\lmed$ under $G_{\rm SM}$ \\
\hline 
\hline 
$\fRR$ & 1/2 & 0 & ${\bf (3,1,2/3)}$ & ${\bf (1,1,-1)}$ \\
\hline 
$\fRL$ & 1/2 & 0 & ${\bf (3,1,2/3)}$ & ${\bf (1,2,-1/2)}$ \\
\hline 
$\sRR$ & 0 & 1/2 & ${\bf (3,1,2/3)}$ & ${\bf (1,1,-1)}$ \\
\hline 
$\sRL$ & 0 & 1/2 & ${\bf (3,1,2/3)}$ & ${\bf (1,2,-1/2)}$ \\
\hline
$\fRRd$ & 1/2 & 0 & ${\bf (3,1,-1/3)}$ & ${\bf (1,1,-1)}$ \\
\hline 
$\fRLd$ & 1/2 & 0 & ${\bf (3,1,-1/3)}$ & ${\bf (1,2,-1/2)}$ \\
\hline 
$\sRRd$ & 0 & 1/2 & ${\bf (3,1,-1/3)}$ & ${\bf (1,1,-1)}$ \\
\hline 
$\sRLd$ & 0 & 1/2 & ${\bf (3,1,-1/3)}$ & ${\bf (1,2,-1/2)}$ \\
\end{tabular}
\end{center}
\caption{The simplified models studied in this paper.
\DM could be either spin 1/2 or 0, which fixes the spin of the colored and uncolored mediators.
We assume that \DM couples to only right-handed quarks, and but couple to either
right-handed or left-handed leptons.
This choice picks the transformations of the mediators under the \SM gauge group $G_{\rm SM} \equiv SU(3)_c \otimes SU(2)_W \otimes U(1)_Y$.
}
\label{tab:fields}
\end{table*}

\section{The Models}
\label{sec:models}

Our study focuses on simplified models in which \DM $\chi$ has renormalizable Yukawa interactions with \SM fermions $f$ through a partner field $\tilde{F}$, with the interaction schematically given by $\mathcal{L} \supset \chi\tilde{F}{f}$.
These are sometimes called ``$t$-channel" simplified models in reference to the $t$-channel exchange of $\tilde{F}$ in \DM annihilation. 
It is usually assumed that a $Z_{2}$ symmetry under which all non-\SM fields are charged odd (and \SM fields charged even) is responsible for \DM stability.

We consider models comprising two \SM singlets $\chi_{A,B}$, motivated by the possibility that their mass parameters may be tuned to interpolate between a limit of self-conjugacy, i.e. \DM is Majorana or real scalar, and a limit where \DM is Dirac or complex scalar.
We also introduce a colored field $\qmed$ to mediate the singlets' interactions with quarks, and an uncolored field one $\lmed$ to mediate their interactions with leptons. 
If $\chi_{A,B}$ are fermions, the mediators $\qmed$ and $\lmed$ are complex scalars, while if $\chi_{A,B}$ are real scalars the mediators are fermions. 
We consider the following interaction Lagrangian involving these fields:
\beq
\nn \Lc \supset - \sqrt{2}(\lambdaq \qmed \chi^\dagger_B q^\dagger + \lambdal \lmed \chi^\dagger_B \ell^\dagger)+ {\rm H.c.}~,
\label{eq:L}
\eeq 
where we have suppressed indices denoting fermion chirality and flavor.
For spin-1/2 \DM, the most general \DM mass Lagrangian is given by
\begin{equation}
\Lc_{\rm mass} =  \left(\chi_A~~\chi_B\right) 
\begin{pmatrix}  \delm ~~~ \mdm \\
    \mdm ~~~ \delm' \end{pmatrix}
  \begin{pmatrix} \chi_A \\ \chi_B \end{pmatrix} + {\rm H.c.}~.
  \label{eq:massfermion}
\end{equation}
A similar-looking (squared) mass matrix may be written down for spin-0 \DM in terms of the field $\phi_\chi \equiv (\chi_A + i \chi_B)/\sqrt{2}$ and its conjugate $\phi_\chi^\dagger \equiv (\chi_A - i \chi_B)/\sqrt{2}$
\begin{equation}
\Lc_{\rm mass} =  \half \left(\phi_\chi~~\phi^\dagger_\chi \right) 
\begin{pmatrix} \delm^2 & ~ \mdm^2 \\
  \mdm^2 &~~ \delm'^2  \end{pmatrix}
  \begin{pmatrix} \phi_\chi \\ \phi^\dagger_\chi \end{pmatrix} + {\rm H.c.}~.
  \label{eq:massscalar}
\end{equation}
The fields $\chi_{A,B}$ mix to give mass eigenstates $\chi_{1,2}$, with the lighter species $\chi_1$ serving as \DM.
In our analysis we will refer to this \DM field as simply $\chi$.
The mediator masses are free parameters that need not originate from symmetry breaking. 
For instance, they may arise from the  scalar potential if the mediator is spin-0, or could be vector-like if \DM is spin-1/2.


\begin{figure*}[t!]
\begin{center}
\includegraphics[scale=0.55]{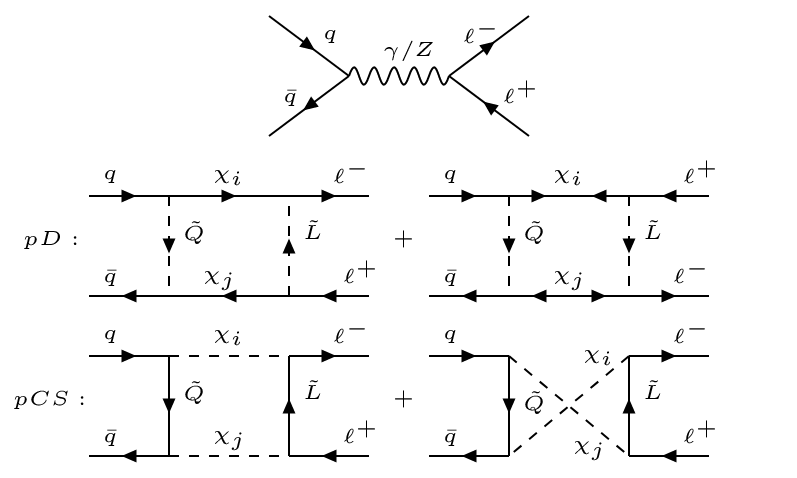}
\end{center}
\caption{
Feynman diagrams for dilepton production at the \LHC.
On top is the Standard Model Drell-Yan process at tree level. 
The middle row shows the box contributions from pseudo-Dirac \DM with scalar mediators.
The bottom row shows the same from pseudo-complex \DM with fermion mediators.
See the text and Table~\ref{tab:fields} for more details.
}
\label{fig:Feyn}
\end{figure*}
 
We now make the following assumptions that simplify our analysis:

\begin{enumerate}
\item We assume a common mass $\mmed$ for the colored and uncolored mediators, and equal \DM couplings to quarks and leptons, $\lambda \equiv \lambdaq = \lambdal$.

\item We assume that \DM couples to only a single chirality of \SM fermions.
This restricts the number of mediator species, since otherwise one would need to introduce mediators that are both singlet and doublet under $SU(2)_W$.
We will consider couplings to both left-handed and right-handed leptons, but only couplings to right-handed quarks.
We do not consider couplings to left-handed quarks because, due to $SU(2)_W$ invariance, they will lead to new physics (\acro{NP}) signals at once from both up-type and down-type quarks in the initial state.
These contributions affect proton-level cross sections in non-trivial ways due to differences in parton densities between up and down quarks, which is a complication we wish to avoid in our analysis.
For further simplicity, we only consider couplings to electrons and muons, and to either the right-handed up quark or the right-handed down quark.
This can be arranged by a special flavor structure, which we spell out next.

\item In order to avoid flavor changing neutral currents, we assume the existence of three generations of mediators with their couplings aligned with the \SM Yukawa couplings such that, in the mass basis, each mediator generation couples only to a single generation of \SM fermions.
In order for \DM to couple solely to the up/down quark, or to the electron/muon, we assume that mediators of the other generations are heavy.

\item As manifest in Eq.~(\ref{eq:L}), we assume that only $\chi_B$ interacts with the \SM fermions.
This assumption captures all the qualitative features of our results;
allowing both $\chi_A$ and $\chi_B$ to interact tends to only rescale the couplings required to produce similar signal rates.

\item Setting $\delm' = 0$ and varying $\delm$, we can interpolate between Majorana and Dirac (or real and complex scalar) scenarios.
Specifically, the Majorana (or real scalar) limit is achieved by tuning $\delm$, with $\delm \ra \infty$  \cite{1307.7197,1411.6743}, while $\delm \ra 0$ renders spin-1/2 \DM Dirac and spin-0 \DM a complex scalar. 
Pure Dirac/complex scalar \DM notoriously has a large spin-independent cross section scattering off nuclei, and is excluded by direct detection experiments for the range of \DM masses and couplings of interest.
Therefore, in our study we will never truly take $\delm \ra 0$, setting $\delm = 1$ MeV as the lower limit.
 As discussed in Ref.~\cite{1411.6743}, for splittings of this size \DM behaves like a Majorana fermion (if spin-1/2, real scalar if spin-0) in direct detection experiments, since the heavier state is kinematically inaccessible given the local \DM velocity $\sim 10^{-3}$.
Majorana/real scalar \DM typically has a much smaller scattering cross section than the Dirac/complex scalar case and hence is much more viable \cite{1307.8120} (see Sec.~\ref{sec:pheno}). Meanwhile, $\mathcal O(\text{MeV})$ mass splitting is well below the \LHC detector resolution, hence $\chi_{1,2}$ are indistinguishable at colliders and \DM will appear as a Dirac or complex scalar particle in our collider study. 
Thus, for a fixed \DM mass, varying $\delm \ge 1$ MeV will have no effect on how \DM appears in direct detection as all scenarios will interact as  Majorana/real scalars. However, as we will see, $\delm$ will dramatically change how \DM appears in dilepton distributions. For the remainder of this paper, we will refer to the $\delm \ge1$ MeV regime as ``pseudo-Dirac" for spin-1/2 \DM and ``pseudo-complex-scalar" for spin-0 \DM.  

\item We assume that \acro{CP}-violating phases in the masses and couplings vanish.

%

\item We neglect quartic couplings involving new scalars introduced in our set-up, as they have little impact on our dilepton signals.
\end{enumerate}

To summarize, we assume that \DM couples to either right-handed up or down quarks and to electrons or muons of either chirality, with \DM itself having spin 0 or 1/2. Thus we may classify our set-up into eight models, which we dub $\fRR, \fRL, \sRR, \sRL, \fRRd, \fRLd, \sRRd$, and $\sRLd$. The superscript denotes the quark to which \DM couples, and the first (second) subscript the chirality of the quark (lepton), while ``{\tt pD}" and ``{\tt pCS}" denote whether \DM is pseudo-Dirac or a pseudo-complex scalar.
The field content of these models is summarized in Table~\ref{tab:fields}. 


\section{Discriminating signals}
\label{sec:signals}

In this section we illustrate the various effects of radiative corrections from the dark sector on $p p \ra \ell^+ \ell^-$ spectra, and how these may help distinguish the properties of $\chi$. We will go about this task by contrasting the signals produced by mutually exclusive cases of a single property, keeping everything else the same, 
e.g. we will compare signals of $\fRR$ and $\sRR$ while keeping all masses and self-conjugation properties the same.

Assuming massless quarks and leptons, and denoting by $\theta$ the centre-of-momentum scattering angle between the incoming quark and outgoing lepton, the parton level leading order (\acro{LO}) Drell-Yan double differential cross section is given by 

\bea
\nn d\sigma_{\rm tot} &\equiv&  \frac{d^2\sigma_{\rm tot}}{d\ct \ d\mll}  \\
 &=& d\sigma_{\rm SM} + d\sigma_{\rm int} + d\sigma_{\rm \chi}~,
\label{eq:XSexpansion}
\eea
with
\bea
\nn d\sigma_{\rm SM} &=& \frac{1}{32 \pi \mll^2 N_c} \sum_{\rm spins} |\mathcal{M_{\rm SM}}|^2~, \\
\nn d\sigma_{\rm int} &=& \frac{1}{32 \pi \mll^2 N_c} \sum_{\rm spins} 2{\rm Re}(\mathcal{M_{\rm SM}}\mathcal{M_\chi^*})~, \\
d\sigma_{\chi} &=& \frac{1}{32 \pi \mll^2 N_c} \sum_{\rm spins} |\mathcal{M_\chi}|^2~,
\label{eq:XSdefs}
\eea

\noindent
where $N_c = 3$ is the number of \QCD colors, $\mathcal{M_{\rm SM}} = \mathcal{M_{\gamma}} + \mathcal{M_{ Z}}$ is the \SM amplitude for the tree-level Feynman diagram in Fig.~\ref{fig:Feyn}. 

As our \acro{NP} effects enter at loop level, care must be taken to ensure that all effects at a given coupling order are consistently included. Additionally, purely \SM loop (mainly \acro{QCD}) effects must be accounted for. 
These issues give rise to the following considerations:

\begin{itemize}

\item One-loop \acro{SM} effects enter at the amplitude level at $\mathcal O(g^2 g^2_s)$, where $g$ and $g_s$ are the \acro{QED} and \QCD couplings, while \acro{NP} effects enter at $\mathcal O(g^2\lambda^2)$ for vertex corrections and $\mathcal O(\lambda^4)$ for the box diagrams. 
The net result of the purely \SM loop effects is to replace $d\sigma_{\rm SM}$ in Eq.~(\ref{eq:XSdefs}) by the \SM cross section at next to leading order (in \acro{QCD}), $d\sigma_{\rm SM, NLO}$.

\item Interference between \SM and \acro{NP} loops ($d\sigma_{\rm int} $) results in contributions to $d\sigma_{\rm tot}$ of $\mathcal O(g^4\lambda^2)$ and $\mathcal O(g^2\lambda^4)$, where the former involve vertex corrections and the latter involve box diagrams.
\footnote{Vertex corrections contain divergent pieces that must be correctly subtracted and subsumed into the renormalization conditions of the theory.} Comparing these terms, we find the box contributions significantly larger when $\lambda \sim$ 1, which is also the regime of couplings where the \acro{NP} effects have enough statistical significance for \LHC bounds to apply.
This happens not only due to the difference in power-counting the couplings, but also because the box diagrams generate more pronounced threshold effects.
Moreover there is a partial cancelation between triangle diagrams with a photon and with a $Z$, as they have opposite signs.

Note also that the \acro{NP} effects do not interfere with the entire \SM amplitude.
As our models involve couplings to a specific set of fermion chiralities, interference proceeds only with the part of the \SM amplitude involving the same set of fermion chiralities.
For instance, the \acro{NP} pieces in $\fRL$ only interfere with $q_R \bar{q}_R \ra Z/\gamma^* \ra \ell_L \bar{\ell}_L$.

\item The $d\sigma_{\rm \chi}$ term involves the square of box and vertex  corrections. 
These are, in principle, the same order in perturbation theory as the interference between the tree-level and \acro{NP} two-loop amplitudes. 
As we have only calculated \acro{NP} effects at one loop, most terms in $d\sigma_{\rm \chi}$ cannot be consistently included in the calculation\footnote{E.g. the cross term between the \acro{NP} vertex correction and the box diagrams is $\mathcal O(g^2 \lambda^6)$, the same as the interference between the tree-level \SM and a \acro{NP} two loop amplitude.}.
An important exception that {\em can} be included consistently is the square of the \acro{NP} box diagrams, which is the only $\mathcal O(\lambda^8)$ contribution to the cross section at any order.

\item As the box diagrams dominate the interference term, in the following discussion we will drop the vertex correction entirely and use ``$\Mc_\chi$" as a loose notation to describe the box amplitude. The resulting cross section expressions $d\sigma_{\rm int}$ and $d\sigma_{\rm \chi}$ are provided in Appendix~\ref{app:formulae}.

\end{itemize}
%

As our focus in this section is on the qualitative differences between various \DM models, rather than between the \SM and \DM, we will work with $d\sigma_{\rm SM, LO}$ for now.
We will return to \NLO \SM effects and the considerations here itemized in Sec.~\ref{sec:pheno}, when we use the dilepton distributions to derive limits.

The most unique feature of $d\sigma_{\rm tot}$  occurs at $\sqrt{\hat{s}} \gsim 2\mchi$, when $\chi$ goes on-shell in the box diagrams in Fig.~\ref{fig:Feyn}, and $\Mc_\chi$ develops an imaginary part ${\rm Im}(\Mc_\chi)$ determined by the optical theorem. According to the optical theorem, ${\rm Im}(\Mc_\chi)$ is proportional to the product of the amplitudes of the tree-level diagrams (with $\chi$'s and fermions as external legs) obtained from ``cutting" the box diagram vertically.
This imaginary part feeds into the real part ${\rm Re}(\Mc_\chi)$ through dispersion relations, causing the amplitude to rapidly rise near the threshold.
At $\sqrt{\hat{s}} \gg 2\mchi$, ${\rm Re}(\Mc_\chi)$ falls away while ${\rm Im}(\Mc_\chi)$ takes over as the dominant contributor to $|\Mc_\chi|^2$.
The net effect of this takeover at $\sqrt{\hat{s}} \gg 2\mchi$ is no more than the addition of a new channel of dilepton production, hence $d\sigma_{\rm tot}$ will be separated from $d\sigma_{\rm SM}$ by some offset. 
All these effects are reviewed in detail in Ref.~\cite{1411.6743}, where the shape of the new physics spectrum was identified as a ``monocline".
(See also Ref.~\cite{Remiddi:1981hn}, which comprehensively reviews dispersion relations.) 
In the following, we show that the above effects also carry the imprint of \DM's microscopic properties, leading to diverse features in dilepton spectra.

\begin{figure}[t!]
\begin{center}
\includegraphics[width=0.45\textwidth]{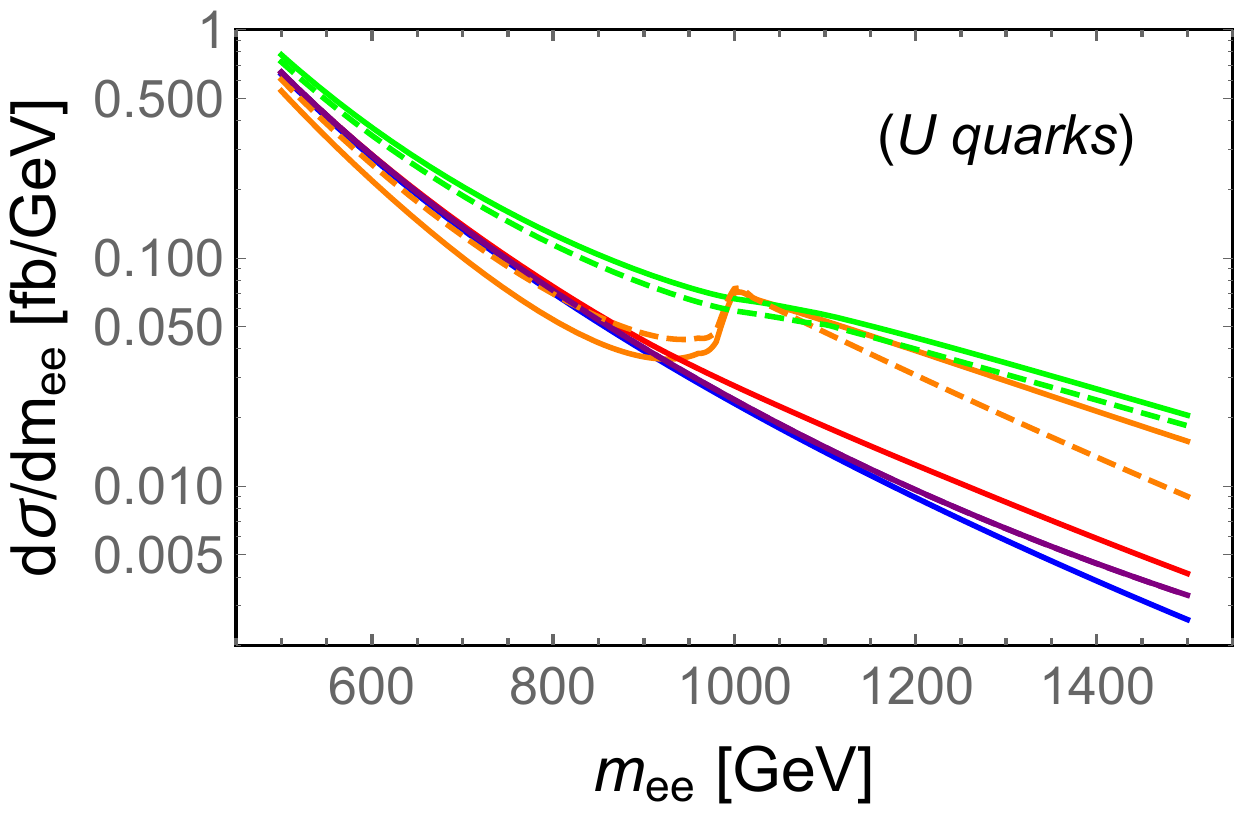}
\end{center}
\caption{Dilepton invariant mass distributions.
The blue line represents the \SM background from Drell-Yan production.
The orange, green, red, and purple represent the pseudo-Dirac, pseudo-complex, Majorana, and real scalar cases. Solid lines represent models with right-handed quarks and right-handed leptons ({\tt RR}) whereas dashed lines represent the {\tt RL} models.
Signal lines are plotted at the benchmark point $\lambda = 2.0, \mdm=500$ GeV, and $\mphi = 550$ GeV. 
}
\label{fig:signalmll}
\end{figure}

For our $\mll$ spectra, we integrate the cross sections in Eq.~\ref{eq:XSexpansion} over $\cos\theta$, and for our angular spectra we integrate them over 400 GeV $\leq \mll \leq$ 4500 GeV, the range used by the 8 TeV \ATLAS analysis \cite{1407.2410}.
The angular spectra are computed in the Collins-Soper reference frame \cite{Collins:1977iv}, in which the directional ambiguity of the initial state quark/antiquark in a $p p$ collider is resolved by boosting to the dilepton center-of-momentum frame and then assuming that the quark originated in the boost direction.
This assumption leads to an $\mll$-dependent probability of initial-quark misidentification, in principle determinable using information of the parton densities (see Appendix A of \cite{1610.03795}).
The scattering angle in this frame is given by
\beq
\ctCS = \frac{Q_{z}}{|Q_{z}|} \frac{2(p_{1}^{+}p_{2}^{-}-p_{1}^{-}p_{2}^{+})}{|Q|\sqrt{Q^{2}+Q_{T}^{2}}}~,
\eeq
 where $Q$ is the net momentum of the dilepton system with $Q_z$ ($Q_T$) the longitudinal (transverse) piece, and $p_{i}^{\pm} \equiv \left(p_{i}^{0}\pm p_{i}^{z}\right)\sqrt{2}$ with
$p_{1}$ ($p_{2}$) the momentum of the lepton (anti-lepton).
Neglecting $Q_{T}$ at high longitudinal momenta, the above may be re-written as
\bea
\nn \ctCS ={\rm sgn}(Q)\tanh\left(\frac{\Delta\eta}{2}\right)~,
\eea
 where $\Delta \eta = \eta_1 - \eta_2$ is the difference in the lepton and anti-lepton pseudo-rapidities.
 
It is often useful to characterize the angular spectrum as a {\em forward-backward asymmetry},
\beq
\afb \equiv \frac{N(\ct > 0) - N(\ct < 0)}{N(\ct > 0) + N(\ct < 0)}~,
\label{eq:afb}
\eeq
or a {\em center-edge asymmetry},
\beq
\ace \equiv \frac{N(|\ct| < \ct_0) - N(|\ct| > \ct_0)}{N(|\ct| < \ct_0) + N(|\ct| > \ct_0)}~,
\label{eq:ace}
\eeq
which marks out how much scattering occurs in central regions.

We now apply the above discussions to our various models.
 All spectra are shown by convolving parton-level cross sections with {\tt MSTW2008NLO} parton distribution functions (PDFs) \cite{0901.0002} at $\sqrt{s} = 13~$TeV.
For our illustrative plots here, we approximate \LHC dilepton production with the Drell-Yan process $q \bar{q} \ra \ell^+ \ell^-$. 
The treatment of secondary processes that also contribute to dilepton production, such as diboson, $t \bar{t}$, dijet and $W$+jets, will become important when we set constraints in Sec.~\ref{sec:pheno}.
We will also show only tree-level \SM cross sections, treating \QCD corrections more carefully in Sec.~\ref{sec:pheno}. 
 
As our current analysis is qualitative, we only show here the behavior of models in which our \DM couples to up quarks, but the broad conclusions we draw hold also for \DM coupling to down quarks. 
See Appendix~\ref{app:downbenchmark} for the signals arising from the latter scenario.
We pick an illustrative benchmark point with the coupling $\lambda$ fixed to 2.0, and masses $\mdm = 500$~GeV and $\mphi = 550~$GeV.
As elaborated in Ref.~\cite{1411.6743}, 
varying $\lambda$ has the effect of raising or lowering $d\sigma_{\rm int}$ and $d\sigma_\chi$.
Depending on the sign of $d\sigma_{\rm int}$, this could enhance or diminish the \DM signal.
Moreover, increasing (decreasing) $\mphi$ enhances (diminishes) the bump feature near $\mll \simeq 2 \mdm$. 
Thus these variations affect the signal significance at the \LHC, a point to which we will return when finding our constraints in Sec.~\ref{sec:pheno}.
Here we note that the spectrum chosen here, being a ``compressed" one, is illustrative of a point where our dilepton probes are expected to outperform jets + $\met$ searches, which suffer from low signal acceptance in these regions.

Finally, in computing our dilepton distributions we impose the following kinematic cuts:
\beq
|\eta_{\ell^{\pm}}| \le 2.4~,~~~~~ p_T^{\ell^{\pm}} \ge 40~{\rm GeV}~.
\label{eq:cuts}
\eeq

We now sketch and contrast the spectral features induced by various \DM species.  
We also elucidate why differences arise between mutually exclusive cases (e.g. spin-0 vs spin-1/2 \DM), and explain how these differences can help us to sort out the properties of \DM and the mediators.
Such a sorting exercise can be successfully carried out at the \LHC if our \DM signals are uncovered with sufficient statistical significance.
Conversely, if the signal-to-background ratios in the event distributions (in our case $d\sigma_{\rm tot}/d\sigma_{\rm SM}$) are inadequate, the \DM properties that can be disentangled could only be a few, or none. 
This may happen if our couplings are small or the mass scales large so as to suppress the effects of new physics amplitudes.
  
\subsection{Self-conjugation}

As explained in Sec~\ref{sec:models}, we may interpolate between the Dirac (complex)  and Majorana (real) limits of \DM by tuning $\delm$.
These limits are readily distinguished by the monocline signature, as shown in Fig.~\ref{fig:signalmll}, where we have plotted $d\sigma_{\rm tot}$ in the non-self-conjugate limit $\delm \ra 0$ for the models $\fRR$ (solid orange), $\fRL$ (dashed orange), $\sRR$ (solid green) and $\sRL$ (dashed green), as well as at the self-conjugate limit $\delm \ra \infty$ for the models $\fRR$ (solid red) and $\sRR$ (solid purple).
In the self-conjugate limit, a subdued signal is produced, while non-self-conjugate \DM can produce large, detectable signals.
Also, in the self-conjugate limit $\fRL$ ($\sRL$) gives near-identical cross sections as $\fRR$ ($\sRL$).
One may compare all these signals with the blue curve, which corresponds to $d\sigma_{\rm SM}$. 

In the Majorana limit, where $\chi_2$ is completely decoupled and only diagrams with $\chi_1$ contribute to the signal, the models $\fRR$ and $\fRL$ produce suppressed signals due to destructive interference between the standard box and crossed box amplitudes.
This arises from a relative minus sign due to an odd permutation of spinors.
As explained in \cite{1411.6743}, this can also be understood in terms of the intricate pattern of interferences between the four standard and four crossed boxes that makes them cancel out one another.
The monocline feature is inferred to appear at $\mdm + (\mdm + \delm) \ra \infty$, so that $d\sigma_{\rm tot}$ remains close to $d\sigma_{\rm SM}$ across $\mll$.

The suppression of rates in the real scalar limit of $\sRR$ and $\sRL$ occurs for subtler reasons.
Due to our choice of coupling to a single fermion chirality, the projection operators pick only the momentum piece in the numerator of the propagator of $\lmed $.
The momentum flow in this propagator in the crossed box diagram is reversed with respect to the standard box (while the fermion flows are the same); consequently, a relative minus sign between the two amplitudes appears, giving rise to the rate suppression.
   
In the limit $\mphi, \mdm \gg \hat{s}$, where the loops can be shrunk to contact operators, the suppressions in the self-conjugate limit are consistent with the loop functions given in the effective theory treatment of Ref.~\cite{1408.1959}.
Since the suppressed rates are a result of a modest addition to $\Mc_{\rm SM}$ from the dark sector in the self-conjugate limit, no sizable signals appear in the angular spectra either.

Finally, we re-emphasize that our ``non-self-conjugate limit" does {\em not} correspond to Dirac or complex scalar \DM, but only to the limit where $\delm$ is small enough to be irresolvable at colliders while remaining large enough  to evade direct detection constraints\footnote{Of course, if the stabilizing $Z_2$ symmetry were broken such that $\chi_1$ decays well within the lifetime of the universe, a pure Dirac or complex scalar formed with $\chi_{1,2}$ is viable.
Then $\chi_1$ is no longer the galactic dark matter searched for at direct detection, and only collider constraints apply.
The decay length of $\chi_1$ determines whether \MET + X or a displaced vertex is the relevant signature.
In all cases our dilepton signatures apply, though the effect of non-trivial widths must now be carefully treated.}.


\begin{figure}[t!]
\begin{center}
\includegraphics[width=0.45\textwidth]{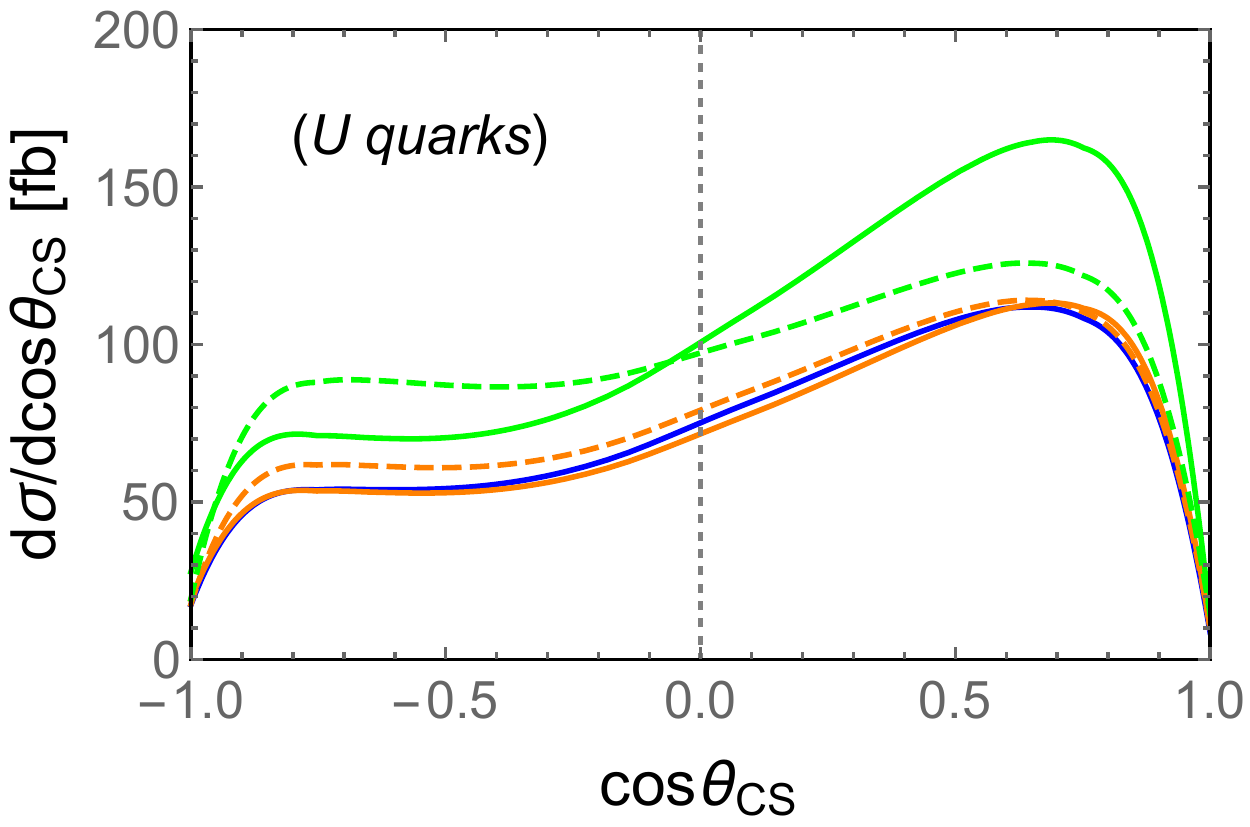}
\end{center}
\caption{Dilepton angular distributions in the Collin-Soper frame. 
The color code and model parameters are as in Fig.~\ref{fig:signalmll}.}
\label{fig:signalcostheta}
\end{figure}

\begin{figure*}[t!]
\begin{center}
\includegraphics[width=0.45\textwidth]{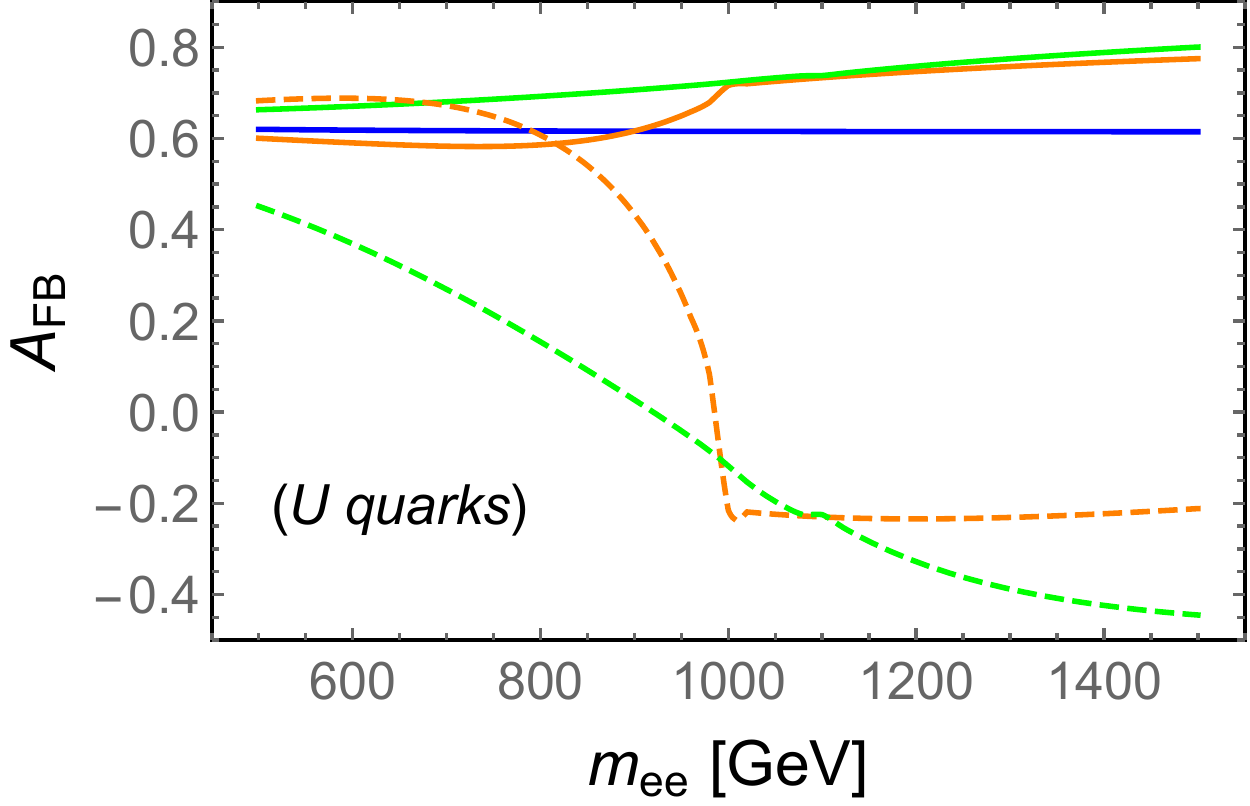} \quad\quad
\includegraphics[width=0.45\textwidth]{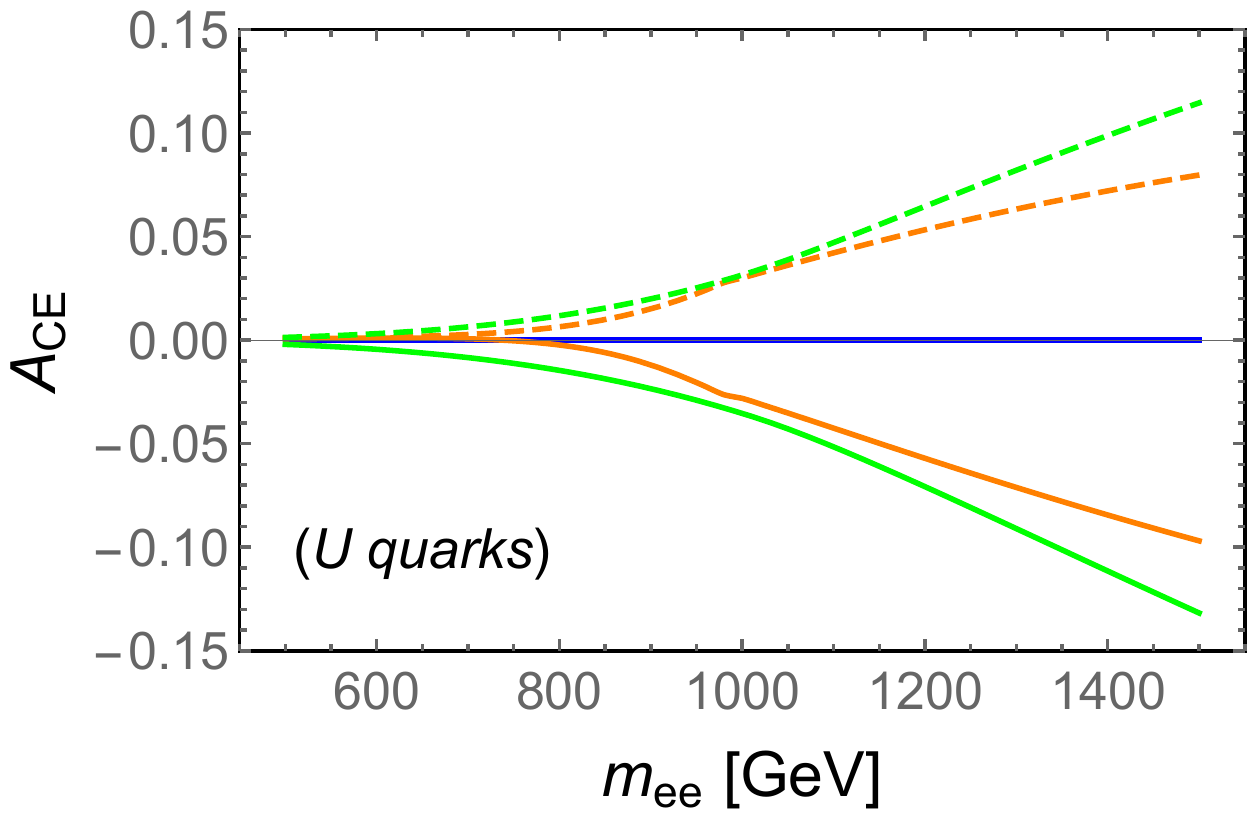}
\end{center}
\caption{Forward-backward asymmetry (left) and center-edge asymmetry (right) at the parton level, as defined in Eqs.~\ref{eq:afb} and \ref{eq:ace}. 
The color code and model parameters are as in Fig.~\ref{fig:signalmll}.}
\label{fig:signalafbace}
\end{figure*}

\subsection{Spin}
Distinguishing the spin of \DM is more challenging than the self-conjugation property, but some headway can be made.
In Fig.~\ref{fig:signalmll}, we see a pronounced ``kick" in the signal rates at $\mll \simeq 2 \mdm$ for fermionic \DM ($\fRR, \fRL$), while the rise in rates appears gentle for scalar \DM ($\sRR, \sRL$).
This may be understood from the fact that near threshold, the box amplitude is determined by Im($\Mc_\chi$), which, as mentioned above, is in turn determined by the tree-level amplitudes for $f\bar{f} \ra \chi \chi$.
The pair-production of complex scalar $\chi$ is more phase-space suppressed at threshold than a Dirac $\chi$, ultimately resulting in a subdued slope of the rise in $d\sigma/d\mee$ for spin-0 \DM.
In any case, even this difference fades for larger mass splittings between the mediators and \DM, where the kick feature is not as pronounced.  

One would naively expect the angular distributions to discern the spin of \DM, on the strength of their ability to clearly distinguish the spin of mediators in the 
$s$-channel \cite{1107.5830} and $t$-channel \cite{1610.03795}.
However, this does not turn out to be entirely true.
While angular spectra are capable of picking up the spin of new particles interfering with the \SM via {\em tree-level} amplitudes, 
the angular spectrum resulting from interference with a {\em loop} amplitude is non-trivial.  
Moreover, as the loop consists of particles with multiple spins, one expects information on the spins to be washed away in the spectrum.
To illustrate this, in Fig.~\ref{fig:signalcostheta} we have shown the angular spectra of our models following the color code of Fig~\ref{fig:signalmll} and using the same benchmark points.
No visible difference in the spectral shape exists between $\fRR$ and $\sRR$; somewhat fewer events populate the $\ctCS < 0$ region for $\fRL$ than for $\sRL$, but this does not amount  to a qualitative difference.
We notice a smaller net deviation from the \SM background for spin-1/2 \DM, which can be understood from their $\mll$ spectra in Fig.~\ref{fig:signalmll}.
Due to negative interference between the tree and box amplitudes, we see a deficit in cross sections with respect to the \SM for $\mll < 2 \mdm$, while an excess appears at $\mll > 2 \mdm$ from the squared box amplitude and threshold effects overwhelming the interference terms.
These deviations are however washed away when integrating over $\mll$, as done for obtaining the $\ctCS$ spectrum.
No such washing away occurs for spin-0 \DM as the tree-box interference is always constructive, giving only an excess of events in the $\mll$ spectrum.
No such washing away would occur for \DM coupling to down quarks either, as the tree-box interference is constructive here as well.
Moreover, the magnitude of the net deviation from background in all cases is sensitive to the $\mll$ window over which cross sections are integrated. 
For these reasons the scattering angle is not a reliable tool to determine the spin of \DM.

\subsection{Mass}

From the previous sub-section, it is apparent that the mass of \DM may be readily cornered if \DM is a fermion and if its mass is not much separated from the mediator's.
In that case, the pronounced kick feature in the $\mll$ signal appears at an invariant mass of 2$\mdm$. 
As this feature is a result of amplitude-level deviations, it must also be reflected in some way in angular observables plotted as a function of $\mll$.
For instance, one would see it in the $\afb$, defined in Eq.~(\ref{eq:afb}), 
plotted at the parton level (for illustration) in the left panel of Fig.~\ref{fig:signalafbace} using the same color code as above.
The behavior of the $\afb$ as a function of $\mll$ with respect to the \SM is in accord with the behavior of the $\mll$ spectrum -- the telltale imprint of interference effects.
Consequently, an abrupt change of slope is visible in the orange curves at $\mll \simeq 2\mdm$.
One would also see the kick feature in $\ace$ (defined in Eq.~\ref{eq:ace})
plotted at the partonic level in the right panel of Fig.~\ref{fig:signalafbace}, where the choice $\ct_0 = 0.596$ sets the \SM value to zero.
Once again the abrupt change of slope at $\mll \simeq 2\mdm$ may be seen in the orange curves. 

In principle, the \DM mass is resolvable for all our models if the mediator mass is of the same order,
a task achievable with sufficiently high statistics, by shape-fitting signals from both $\mll$ and $\ctCS$ spectra to various hypotheses.

\subsection{Chirality}

The relative chirality between the quarks and leptons in the new physics amplitude, i.e. whether the model is {\tt RR} or {\tt RL}, shows up in dilepton spectra in quite interesting ways.
({\tt RR} and {\tt LL}, and separately {\tt RL} and {\tt LR}, yield similar spectra.)
We see a difference in the $\fRL$ and $\fRR$ signals in the $\mll$ spectrum in Fig.~\ref{fig:signalmll}, though both exhibit similar shapes.
This is due to the difference in projection operators in the fermion chains:
in $\fRL$ ($\fRR$), the combination of $P_R$ and $P_L$ ($P_R$ and $P_R$) picks the mass (momentum) piece from the numerator of the $\chi_1$ propagator.
When we turn to the $\sRL$ and $\sRR$ signals, however, we find negligible difference. 
This is because the fermion chains are now different, always coming with the combination $P_R$ and $P_L$ (that now picks the momentum piece from the numerator of the mediator propagators).

Much more revealing differences appear in the angular spectrum; 
we know from $Z$ boson physics and from contact operator analyses (such as in \cite{1407.2410}) that the chiral nature of new states has an impact on the scattering angle. 
Such an impact is seen in the cases where the \acro{NP} amplitude interferes constructively with the tree-level one, i.e. in $\sRL$ and $\sRR$ (also in $\fRLd$ and $\fRRd$ in Appendix~\ref{app:downbenchmark}).
In Fig.~\ref{fig:signalcostheta},
more forward ($\ctCS > 0$) events are produced by $\sRR$ versus $\sRL$, and $\sRL$ produces visibly more backward ($\ctCS < 0$) events than $\sRR$. 
These differences are best seen by plotting the $\afb$, as in the left panel of Fig.~\ref{fig:signalafbace}.
The $\afb$ neatly separates the cases of {\tt RR} and {\tt RL}, putting them above and below the \SM value at large $\mll$. 
Further, the $\ace$ in the right panel of Fig.~\ref{fig:signalafbace} clearly signals the chirality combination by putting {\tt RR} ({\tt RL}) above (below) the \SM value. 
\\

To summarize this section, we have shown that dilepton measurements can carry information on \DM's self-conjugation, spin and mass, and the chiral structure of \DM's interactions with \SM fermions. 
We did this by picking a benchmark point with an interaction strength large enough and \DM/ mediator mass scale small enough to produce clear signal features in dilepton spectra.
Whether these features can be actually discerned at the \LHC will depend on the viability and signal significance of each point in parameter space.
Thus, in the next section we will derive constraints on our models from available \LHC data and project our sensitivity at the high luminosities of the 13 TeV run.
In Sec.~\ref{sec:disc} we will point out interesting regions that are currently viable and can be probed by the future \LHC, which is indicative of regions where the above analysis would apply.

\begin{figure*}[t!]
\begin{center}
\includegraphics[width=0.45\textwidth]{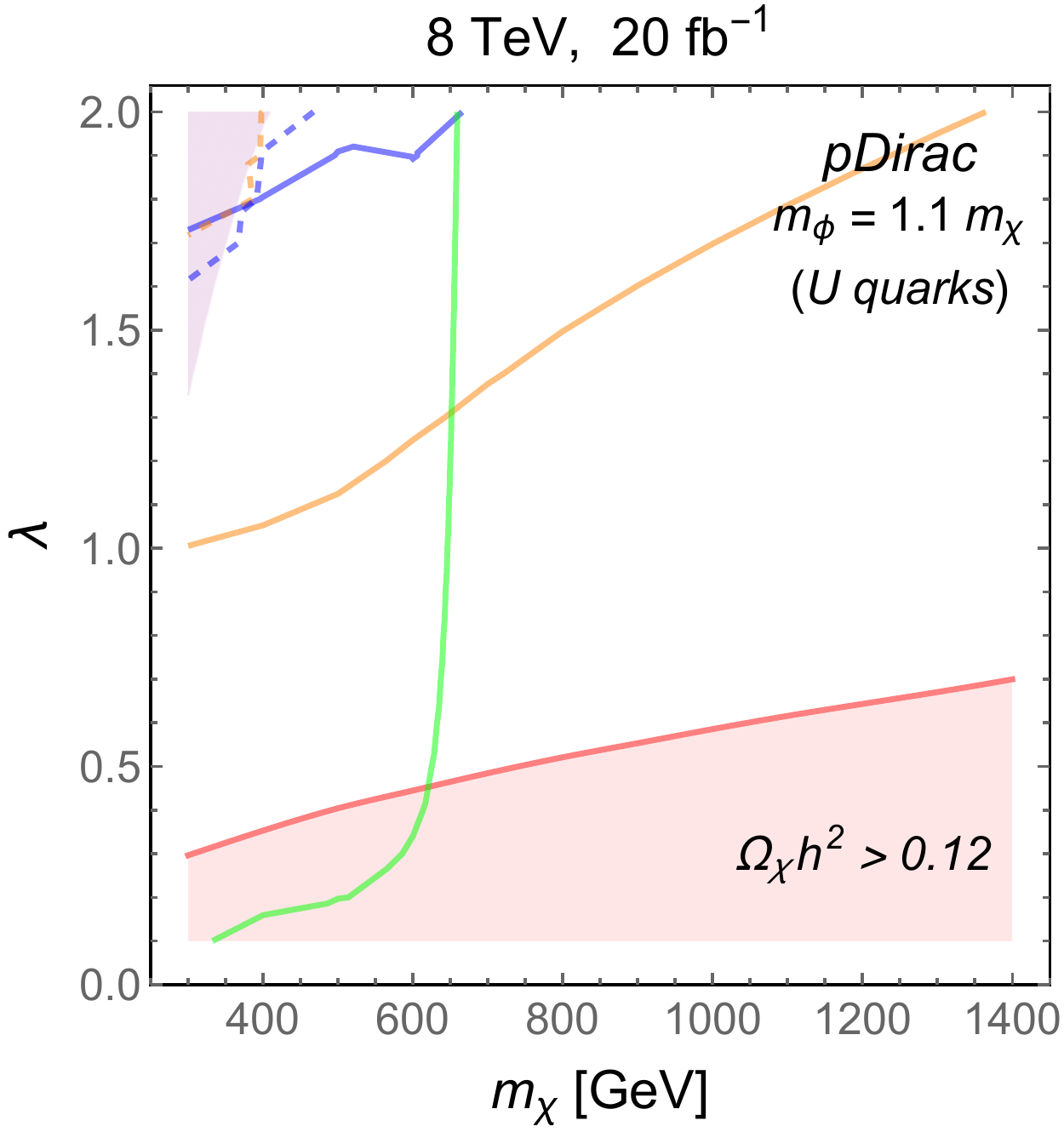}
\quad \quad
\includegraphics[width=0.45\textwidth]{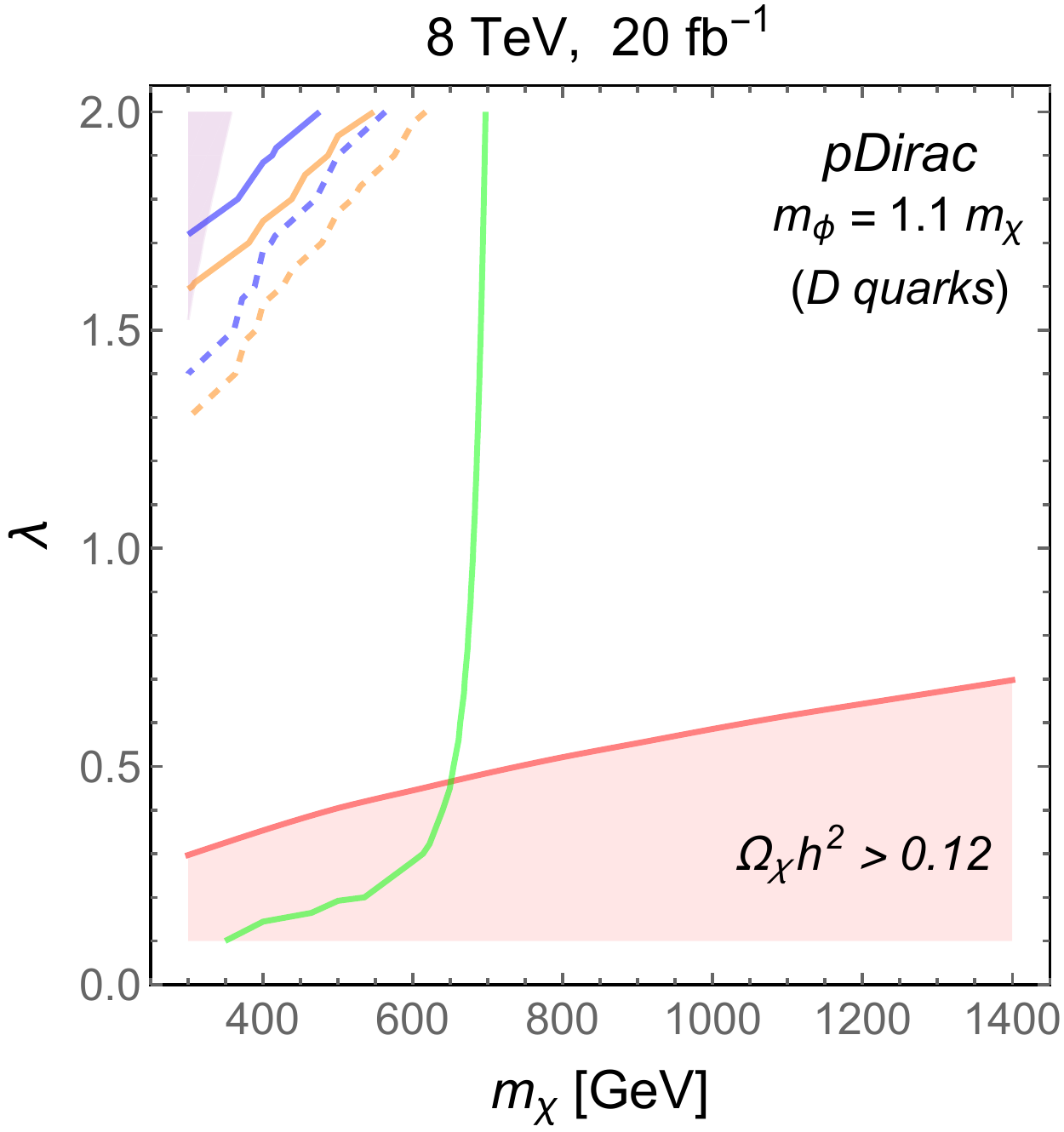}
\\ 
~\\
\includegraphics[width=0.45\textwidth]{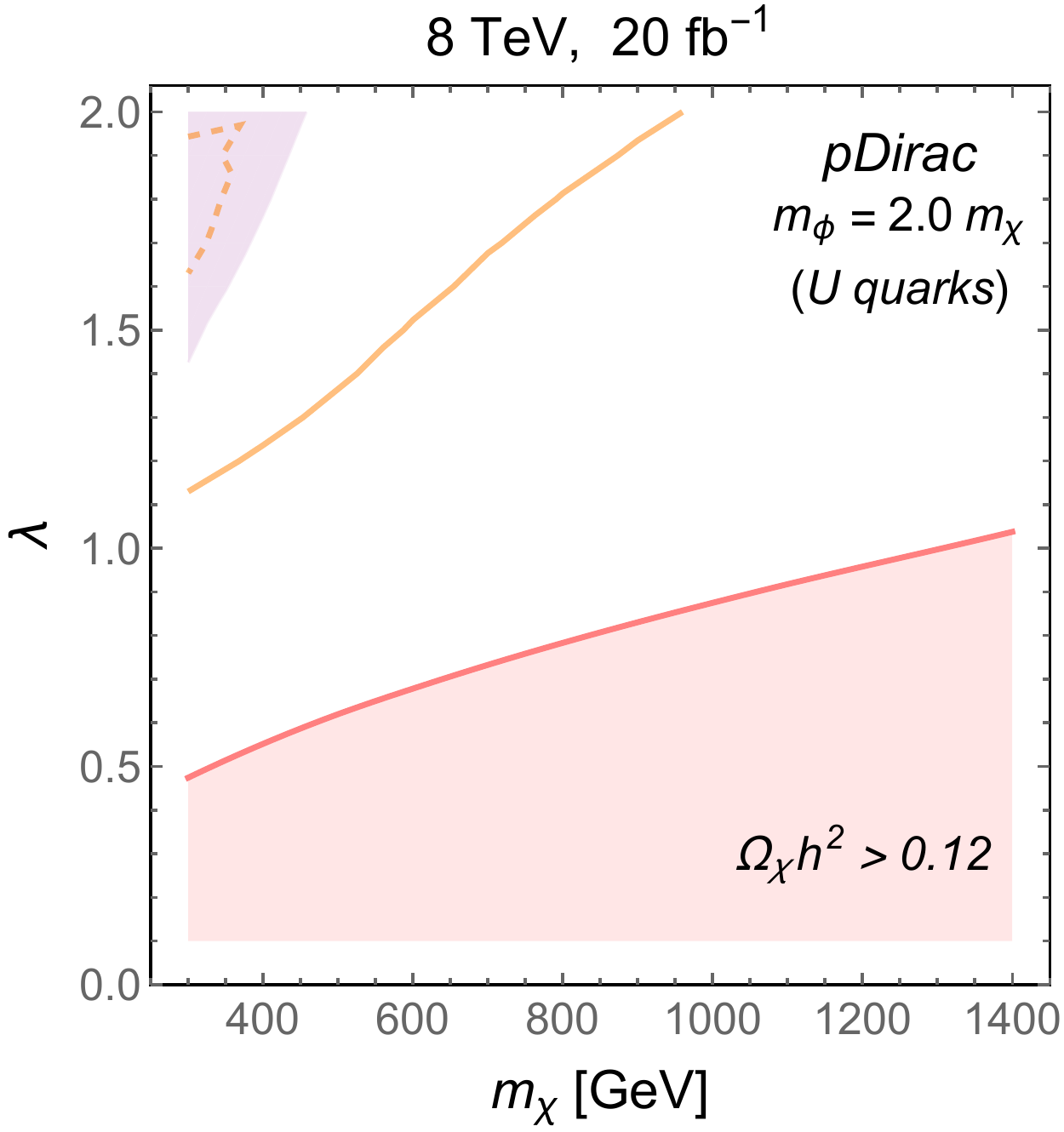}
\quad \quad
\includegraphics[width=0.45\textwidth]{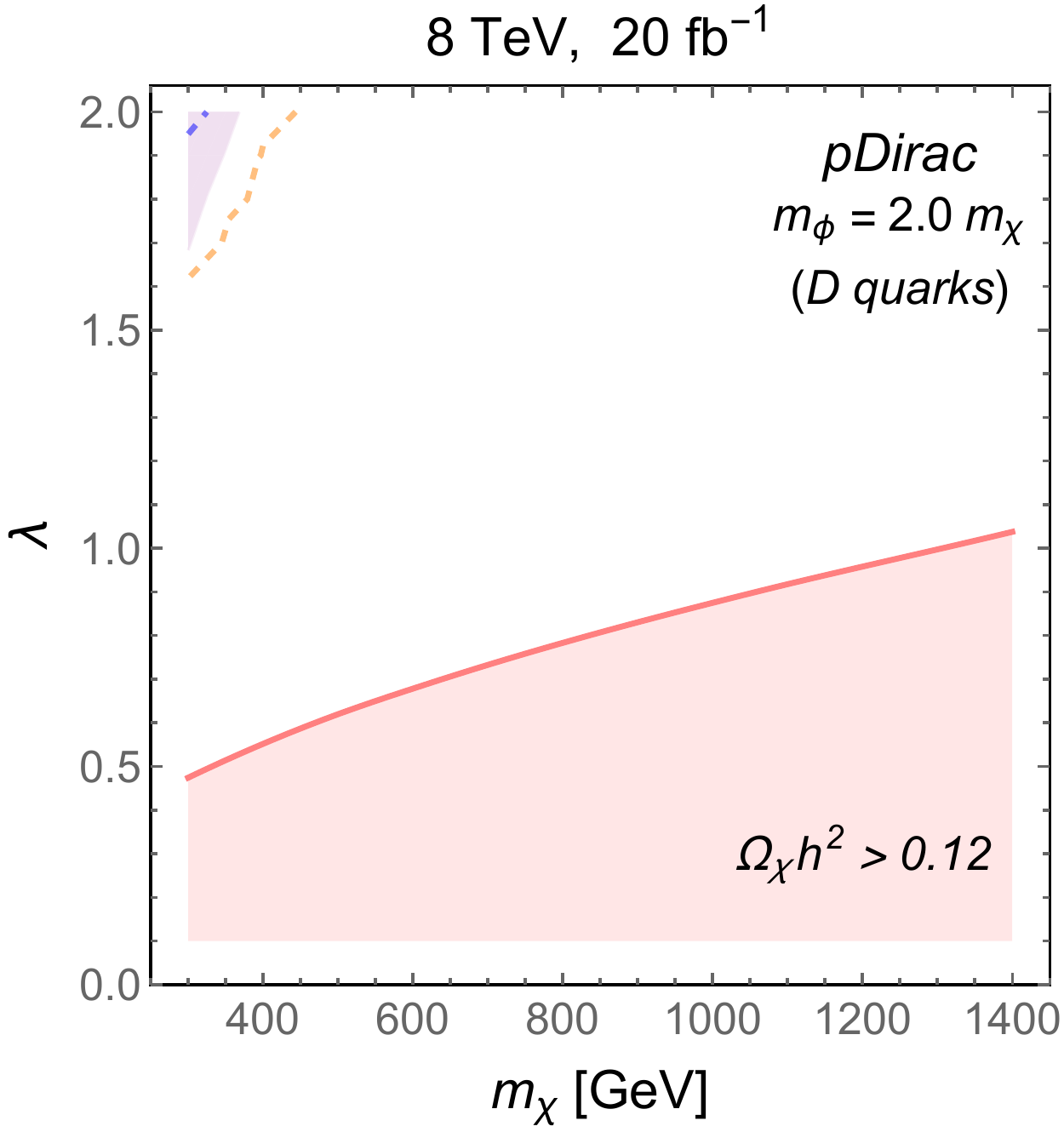}
\end{center}
\caption{
The bounds on our models at two different hierarchies between $\mphi$ and $\mdm$.
Bounds at 95\% \CL from the \LHC are obtained from measurements at 8 TeV and 20 fb$^{-1}$;
the orange (blue) curves depict dilepton bounds on the {\tt RR} ({\tt RL}) models, and are solid (dashed) for $\mee$ ($\ctCS$) bounds; 
the purple regions are excluded by jets + \MET searches.
The green curves are 90\% \CL Xenon1T constraints on spin-independent scattering,
and the red region leads to \DM overabundance through freeze-out.
See text for further details. 
}
\label{fig:bounds}
\end{figure*}

\begin{figure*}[t!]
\begin{center}
\includegraphics[width=0.45\textwidth]{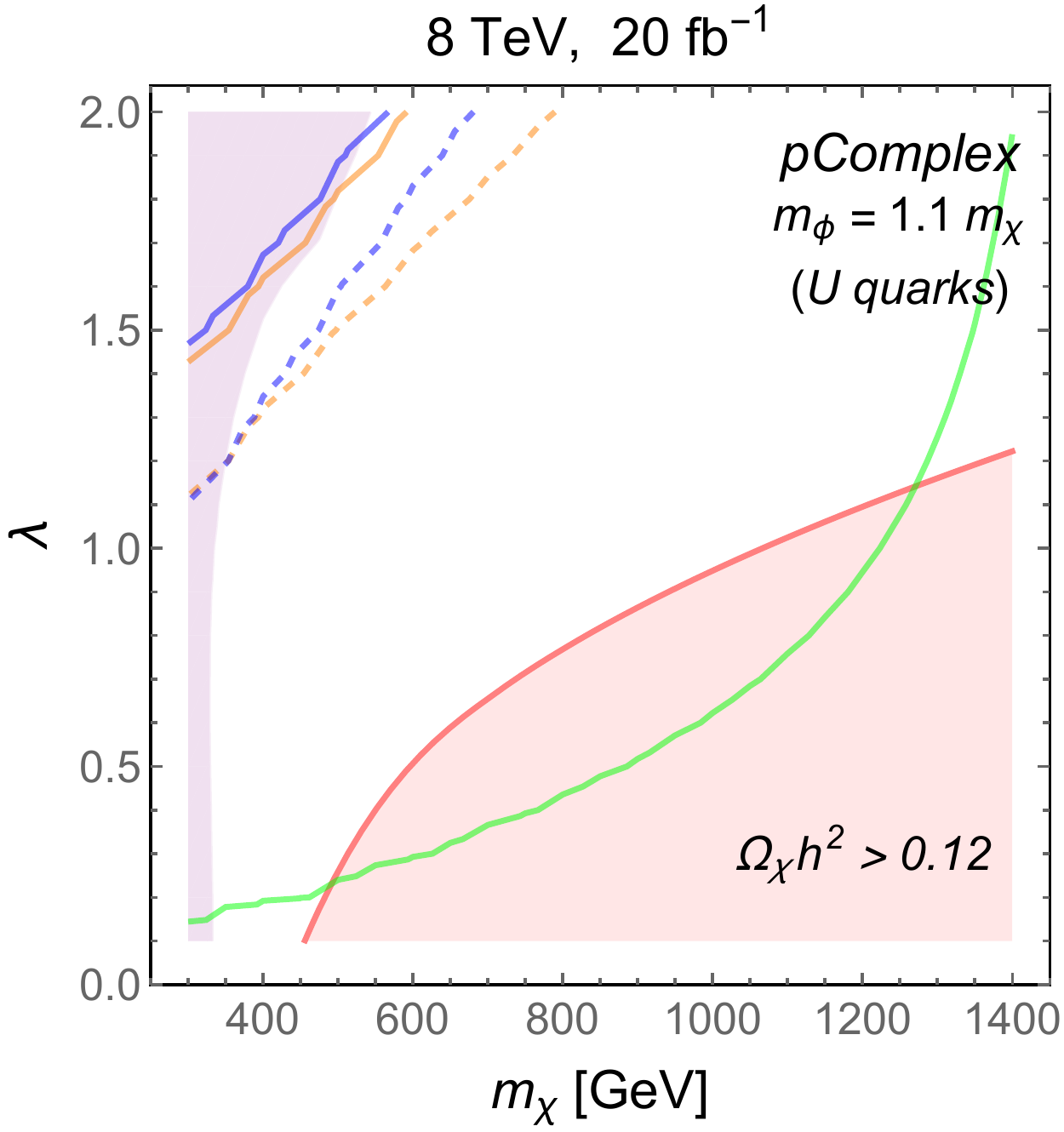}
\quad \quad
\includegraphics[width=0.45\textwidth]{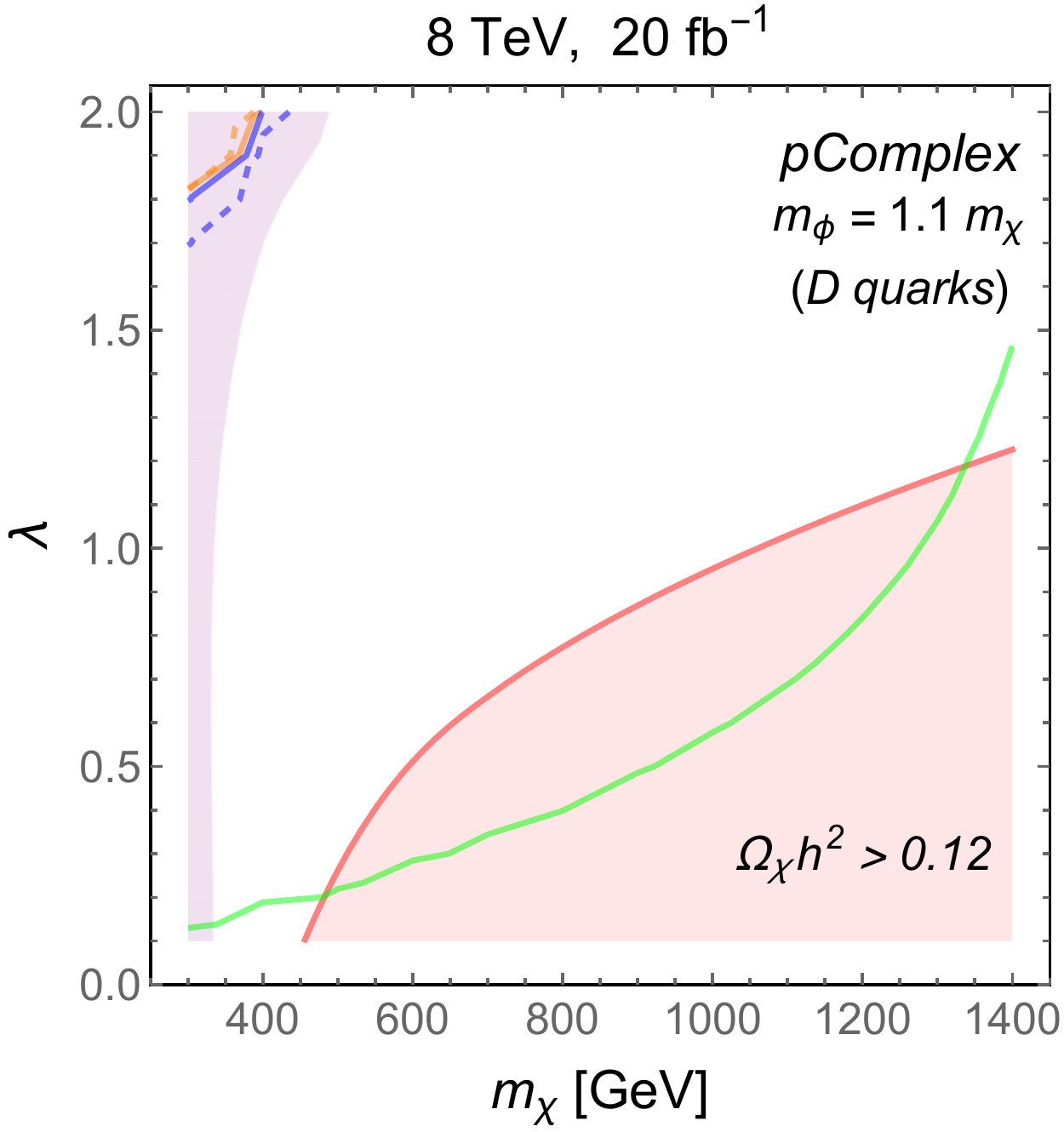}
\\ 
~\\
\includegraphics[width=0.45\textwidth]{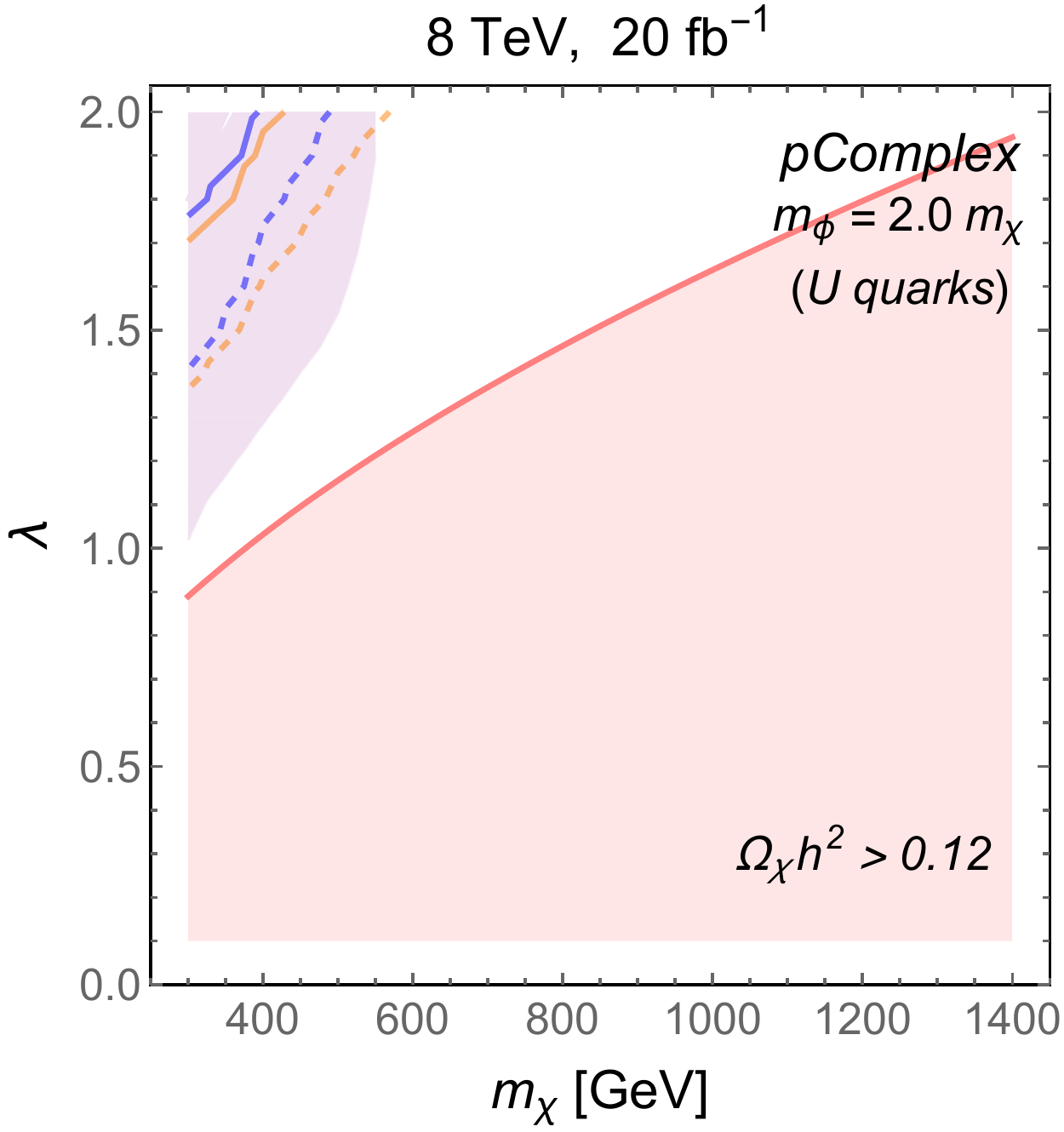}
\quad \quad
\includegraphics[width=0.45\textwidth]{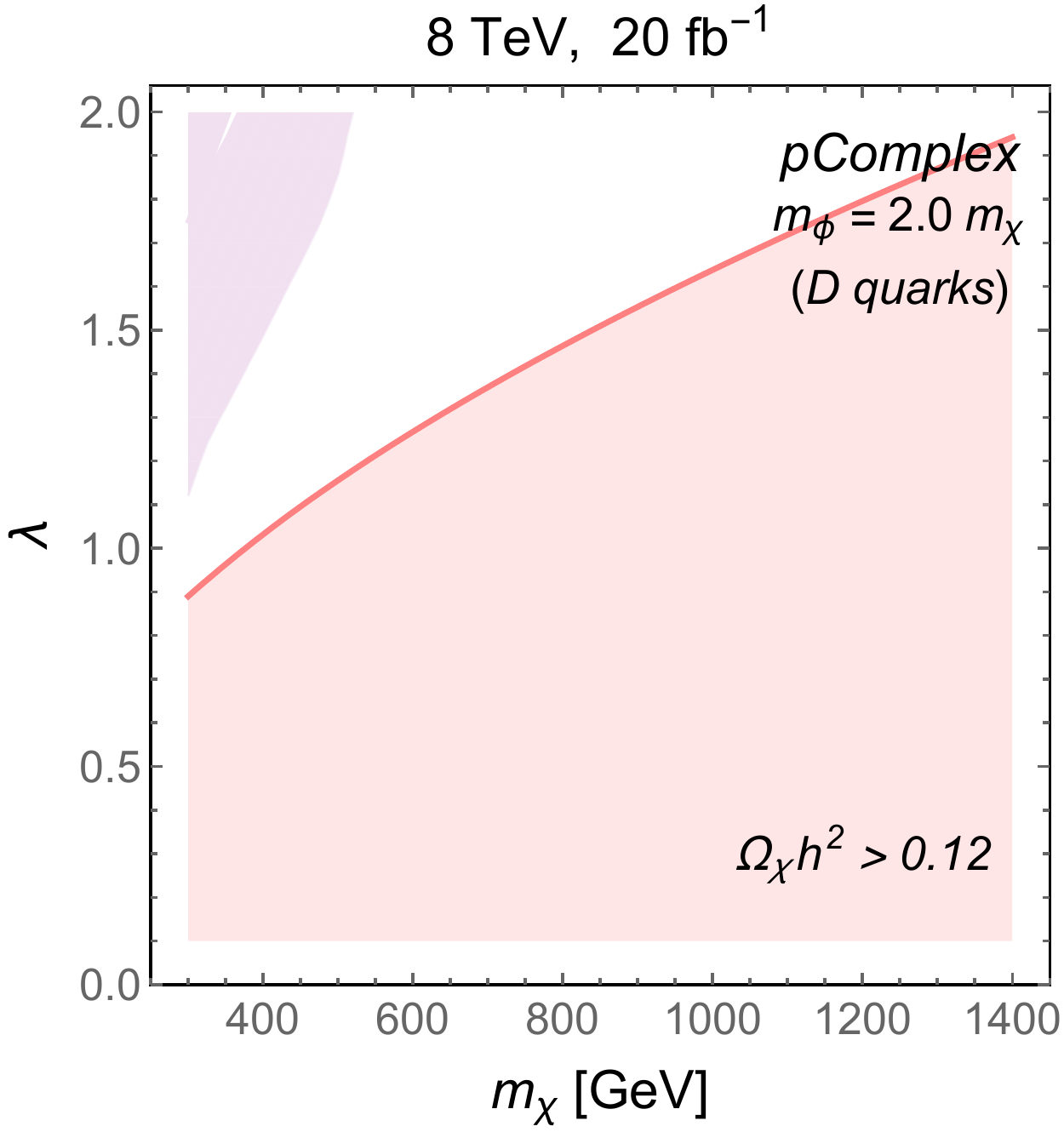}
\end{center}
\caption{
The bounds on our models at two different hierarchies between $\mphi$ and $\mdm$.
Bounds at 95\% \CL from the \LHC are obtained from measurements at 8 TeV and 20 fb$^{-1}$;
the orange (blue) curves depict dilepton bounds on the {\tt RR} ({\tt RL}) models, and are solid (dashed) for $\mee$ ($\ctCS$) bounds; 
the magenta curves depict jets + \MET constraints.
The green curves are 90\% \CL Xenon1T constraints on spin-independent scattering,
and the red region leads to \DM overabundance through freeze-out.
See text for further details. 
}
\label{fig:boundsdown}
\end{figure*}

\section{Constraints and Prospects}
\label{sec:pheno}

Having illustrated that dilepton distributions may help distinguish the properties of \DM, we now show that \DM could in fact reveal itself first in \LHC measurements of dilepton events. 
In this section we will derive constraints on our models from the available \LHC data on $\mee$ and $\ctCS$, and compare them to constraints from conventional \DM searches such as jets + \MET, direct detection, and relic density measurements.
In Appendix~\ref{app:otherconstraints} we discuss a few other probes that set much weaker constraints.
We will find that current $\ctCS$ measurements can outdo all other bounds; this is one of the main results of our paper.

We show the above limits in the plane of the Yukawa coupling $\lambda$ versus \DM mass $\mdm$ in Figs~\ref{fig:bounds} and \ref{fig:boundsdown}, which correspond respectively to spin-1/2 and spin-0 \DM;
the left-hand (right-hand) panels depict \DM coupling to up (down) quarks.
As mentioned in the previous section, these limits must depend on the hierarchy of mediator and \DM masses. 
Thus we pick two benchmark spectra for illustrating the constraints, one where the spectrum is ``compressed" with $\mphi= 1.1 \ \mdm$, and one where it is ``uncompressed" with $\mphi= 2 \ \mdm$. 
These correspond to the top and bottom row respectively.
Throughout our analysis here we fix $\delm$ = 1 MeV.
As explained in Sec.~\ref{sec:models}, this gives our \DM safety from direct detection constraints while masquerading as a Dirac/ complex scalar particle at the \LHC.

We begin our discussion with dilepton constraints.
The orange and blue curves in Figs~\ref{fig:bounds} and \ref{fig:boundsdown} show the 95\% \CL limits from the \LHC Run 1 (8 TeV, 20 fb$^{-1}$) measurements of dilepton spectra, corresponding to the {\tt RR} and {\tt RL} models respectively.
The solid (dashed) curves correspond to $\mee$ ($\ctCS$) measurements by \ATLAS \cite{1407.2410,1405.4123} in $e^+ e^-$ production.
Due to the similarity of results, we expect similar limits from \CMS data \cite{1412.6302,1601.04768} and from dimuon production.
We perform a $\Delta \chi^2$ fit as done in \cite{1411.6743}, but considerably improve on the treatment to obtain realistic bounds.

Broadly speaking, the recasting of dilepton measurements into bounds is performed by comparing between three sets of events across $\mll$ or $\ctCS$ bins (labelled by $i$): the data $N_{d_i}$, the background $N_{b_i}$, and the signal $N_{s_i}$.
We take $N_{d_i}$ from \ATLAS \cite{1405.4123,1407.2410}.
It is useful to divide the background into its dominant and subdominant components, $N_{b_i} = N^{\rm dom}_{b_i} + N^{\rm sub}_{b_i}$. The former comprises of the Drell-Yan $s$-channel process in Fig.~\ref{fig:Feyn}, while the latter (which we also take from \ATLAS) comprises of the reducible backgrounds of diboson, top, dijet, and $W$ + jets production.

To obtain an $N^{\rm dom}_{b_i}$ that is as accurate as possible, we impose the cuts described in Sec~\ref{sec:signals}, and obtain Drell-Yan events at \NLO-\QCD (i.e. at $\Oc (\alpha_s)$) using {\tt MCFM 8.0}  \cite{1605.08011} with {\tt MSTW2008NLO} PDFs \cite{1007.2241}, and a renormalization and factorization scale of $\mll$.
Then we account for the efficiency of lepton reconstruction by scaling our events by a global factor that best matches the Drell-Yan background provided by \ATLAS.
At $\sqrt{s} =8$~TeV, this factor is 0.74 (0.67) for the $\mee$ ($\ctCS$) distribution.

Obtaining the signal events is a subtler process.
First, we obtain the parton level total cross section $d\sigma_{{\rm tot}_i}$ defined in Eq.~(\ref{eq:XSexpansion}).
For reasons explained in Sec.~\ref{sec:signals}, we neglect two terms: the interference between the \SM $\Oc(\alpha_s)$ corrections and $\Mc_\chi$, and all terms involving triangle diagrams.
Next, we convolve $d\sigma_{{\rm SM}_i}$ and $d\sigma_{{\rm tot}_i}$ with {\tt MSTW2008NLO} PDFs to obtain the hadron-level cross sections $d\tilde{\sigma}_{{\rm SM}_i}$ and $d\tilde{\sigma}_{{\rm tot}_i}$.
The $N_{s_i}$ are now obtained by first scaling the dominant background by $d\tilde{\sigma}_{\rm tot}/d\tilde{\sigma}_{\rm SM}$, and then adding the result to the subdominant background:
\bea
\nn N_{s_i} =   N^{\rm dom}_{b_i} \left(\frac{d\tilde{\sigma}_{{\rm tot}_i}}{d\tilde{\sigma}_{{\rm SM}_i}}\right) + N^{\rm sub}_{b_i}~.
\eea

Using all the above information, we compute
\bea
\nn \chi^2_s &=& \sum^{N_{\rm bins}}_{i = 1} \frac{(N_{d_i}-N_{s_i})^2}{N_{s_i} + \delta^2_{{\rm sys}_i}}~, \\
\chi^2_b &=& \sum^{N_{\rm bins}}_{i = 1} \frac{(N_{d_i}-N_{b_i})^2}{N_{b_i} + \delta^2_{{\rm sys}_i}}~,
\label{eq:chisq}
\eea
and locate the 95\% \CL bound at $\Delta \chi^2 \equiv \chi^2_s - \chi^2_b = 5.99$.
Here the systematic errors $\delta_{{\rm sys}_i}$ are taken from \cite{1405.4123,1407.2410}.

Our central findings are best understood by directly comparing the right- and left-hand panels of Figs~\ref{fig:bounds} and \ref{fig:boundsdown}.
The relative behavior of these bounds is dictated by two ingredients -- (i) the PDFs: as the up quark has higher parton densities in the proton than the down quark, one expects stronger dilepton bounds for \DM coupling to up quarks for \DM coupling to down quarks, and (ii) interference effects, or more precisely, the signal contribution of the interference versus the squared box, i.e. $d\sigma_{\rm int}$ versus $d\sigma_{\chi}$ in Eq.~(\ref{eq:XSdefs}).
For example, for the models $\fRR$ and $\fRL$, the tree-level and box diagrams interfere destructively, resulting in a {\em deficit} of events with respect to the \SM for $\mll < 2\mdm$; this may be seen in Fig~\ref{fig:signalmll}.
(On the other hand, the relative sign of the down quark's electric charge with respect to the up quark ensures that tree-box interference in the case of \DM coupling to down quarks is constructive.)   
As this deficit occurs in the low $\mee$ bins, where the event population is high, its contribution to the signal $\chi^2$ is considerable\footnote{Ref.~\cite{1411.6743} had incorrectly flipped the sign of $d\sigma_{\rm int}$ for the model $\fRR$, and had derived a bound weaker than that in this work.}.
This explains why the $\mee$ bound for $\fRR$ (and to some extent $\fRL$) is so much stronger on the left-hand than on the right-hand panels of Fig.~\ref{fig:bounds}. 
At the same time, the $\ctCS$ bounds do not show this hierarchy since the effects of the deficit below and excess above $\mee \simeq 2\mdm$ are washed out by the integration over $\mee$ bins. 

In all four plots, we find the {\tt RR} models more constrained than the {\tt RL} models.
In the spin-1/2 \DM models we understand this from the observation made in Sec.~\ref{sec:signals}, that ${\tt \fRL}$ gives smaller cross sections than ${\tt \fRR}$ due to differences in how interference proceeds between the standard and crossed boxes.
As for the spin-0 \DM models, we see from Fig.~\ref{fig:signalmll} that $\sRR$ yields slightly larger cross sections than $\sRL$ and is thus subject to slightly stronger constraints.
Also, as discussed in Sec.~\ref{sec:signals}, our dilepton signal rates decline with $\mphi/\mdm$ due to propagator suppression in the loop.
This results in the weaker limits in the $\mphi = 2 \mdm$ plots in comparison to the $\mphi = 1.1 \ \mdm$ plots: in fact, the $\fRL$ limits are so weak as to disappear from the parametric range displayed.

We now compare our dilepton results with conventional \DM probes.
In addition to modifying dilepton spectra, our models also have the following 
effects.
\begin{itemize}
\item[(a)] They can pair-produce colored mediators both through \QCD and through exchanging  $\chi$ in the $t$-channel of a $q\bar{q}$-initiated process, and these mediators can  decay to a quark and \DM.
Thus, our models confront constraints from dedicated searches for the mediators 
using jets plus missing energy signatures,
\item[(b)] \DM can annihilate into quarks and leptons through $t$-channel exchange of 
 mediators and freeze out in the early universe, confronting the relic density 
 measurement by Planck,
 \item[(c)] \DM can scatter against nucleons through $s$-channel exchange of the mediator 
 $\qmed$, confronting underground direct detection searches.
 \end{itemize}
 
 Bounds derived as a result of (a) - (c) can be seen in Figs~\ref{fig:bounds} and \ref{fig:boundsdown} along with the dilepton bounds explained earlier. 
 
Turning first to the jets + $\slashed{E}_T$ bounds, the purple regions are excluded at 95\% \CL by the \CMS Run 1 search \cite{1402.4770}.
To determine this bound, we reinterpreted the {\tt T2qq} bounds that assumes squark production through \QCD (that is, the gluino is decoupled) followed by prompt decay to light quark + \LSP.
Specifically, we generated leading-order cross sections for $\qmed$ pair production using {\tt MadGraph5} \cite{1405.0301} and {\tt CTEQ6L1} parton distribution functions \cite{Pumplin:2002vw}, and matched them with exclusion cross sections provided by \CMS.
We assume here that the detector acceptances of our models are similar to the {\tt T2qq} model of \CMS.
Since this constraint is agnostic to the chirality of the lepton in our models (with both spin-0 and spin-1/2 \DM), we do not distinguish between {\tt RR} and {\tt LR}. 

The red curve in the plots corresponds to the thermal line where $\Omega_\chi h^2 = 0.12$, with \DM being overproduced in the red shaded region below.
This curve was obtained using {\tt MicrOmegas}4.3 \cite{1305.0237},
and takes into account co-annihilation between \DM and mediators, which becomes important in the compressed region $\mphi \lsim 1.1 \ \mdm$
\footnote{Co-annihilation also occurs between the eigenstates $\chi_1$ and $\chi_2$, but as their mass splitting $\delm$ = 1 MeV $\ll \mdm$, this practically amounts to the self-annihilation of Dirac/complex scalar \DM.}.
Leptonic modes constitute only a small fraction of the annihilation cross section $\langle \sigma v \rangle$, as opposed to quark modes that come with a color factor of 3.
Therefore, though the {\tt RL} models must give a slightly higher $\langle \sigma v \rangle$ than the {\tt RR} models due to neutrino final states, there is no visible difference in the red curves.
It may also be seen that, barring a non-standard thermal history of the universe, the \DM in our models make up a fraction of the total \DM population in regions where most of our dileptonic bounds apply.

Finally, the green curves show 90\% \CL bounds from spin-independent scattering at Xenon1T \cite{1705.06655}. 
(The current spin-dependent limits are consistently weaker and are not shown.)
To obtain these bounds, we assume that the density fraction of \DM at freeze-out equals the density fraction in the galactic halo today, {\em i.e.} $\Omega_\chi h^2/(0.12) = \rho_\chi/ (0.3 \ {\rm GeV}~{\rm cm}^{-3})$, which effectively scales the exclusion cross sections by $0.12/\Omega_\chi h^2$. 
Our annihilation and scattering cross sections are provided in Appendix~\ref{app:formulae}.

\begin{figure*}[t!]
\begin{center}
\includegraphics[width=0.45\textwidth]{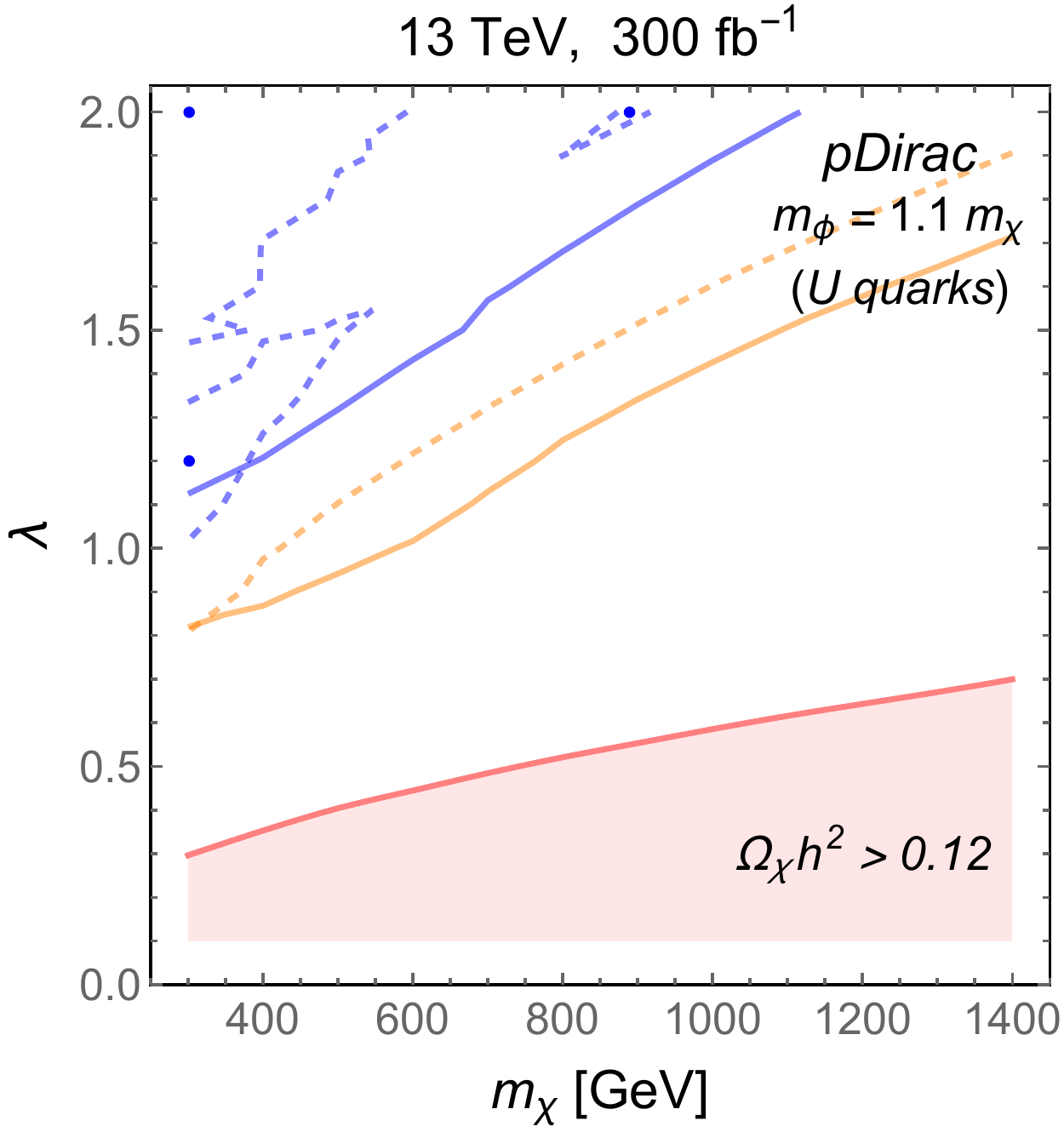}
\quad \quad
\includegraphics[width=0.45\textwidth]{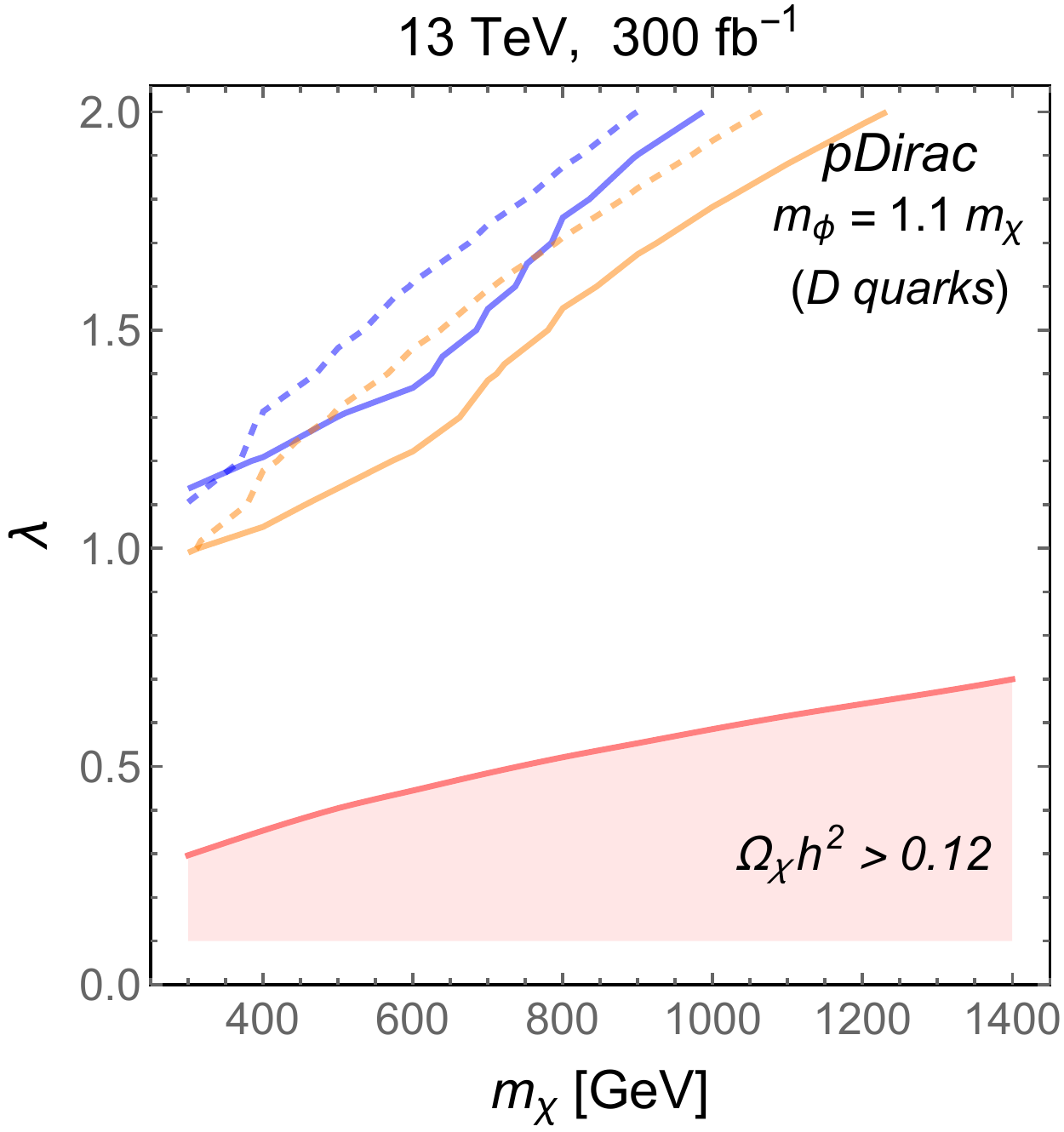}
\\ 
~\\
\includegraphics[width=0.45\textwidth]{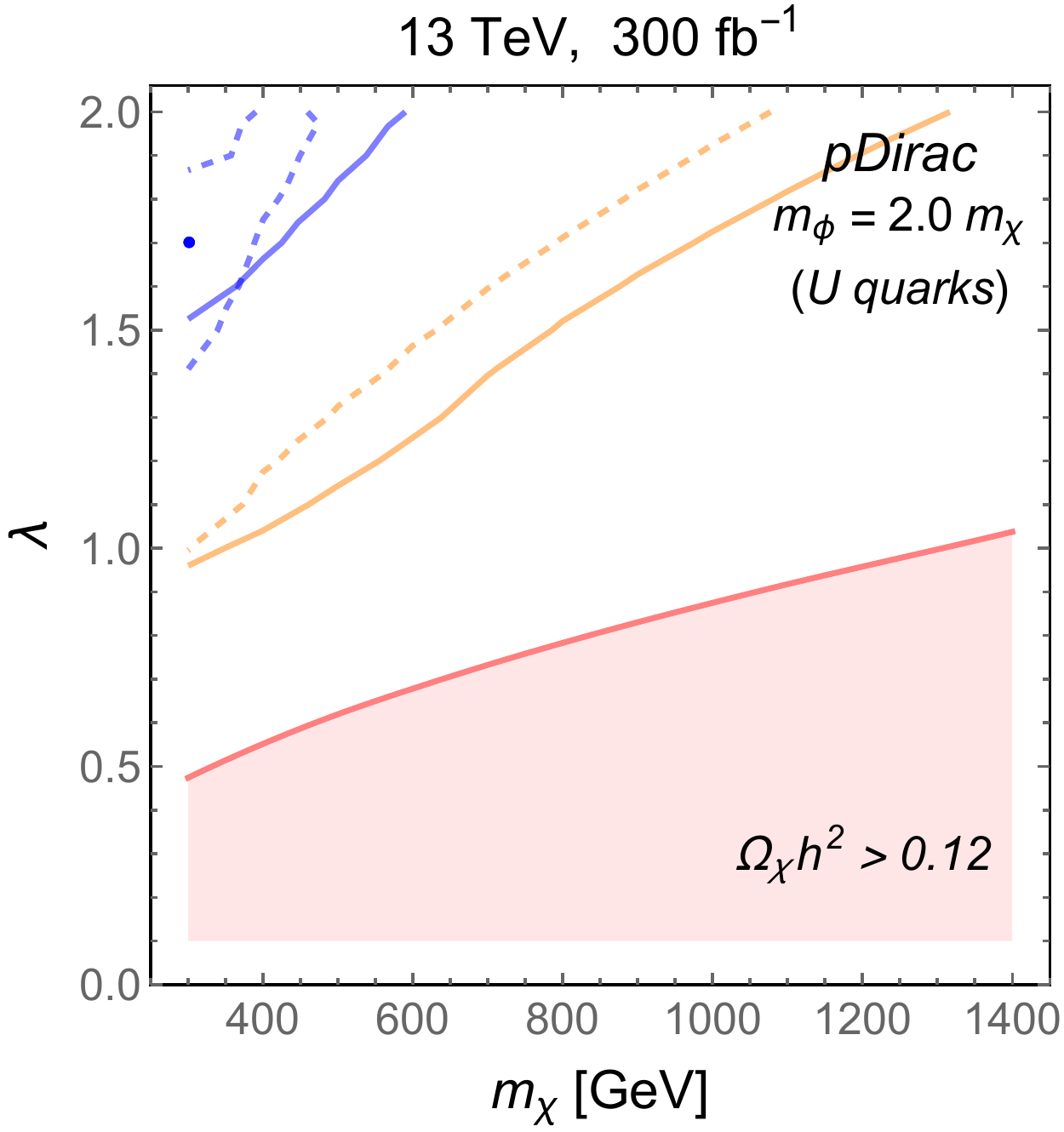}
\quad \quad
\includegraphics[width=0.45\textwidth]{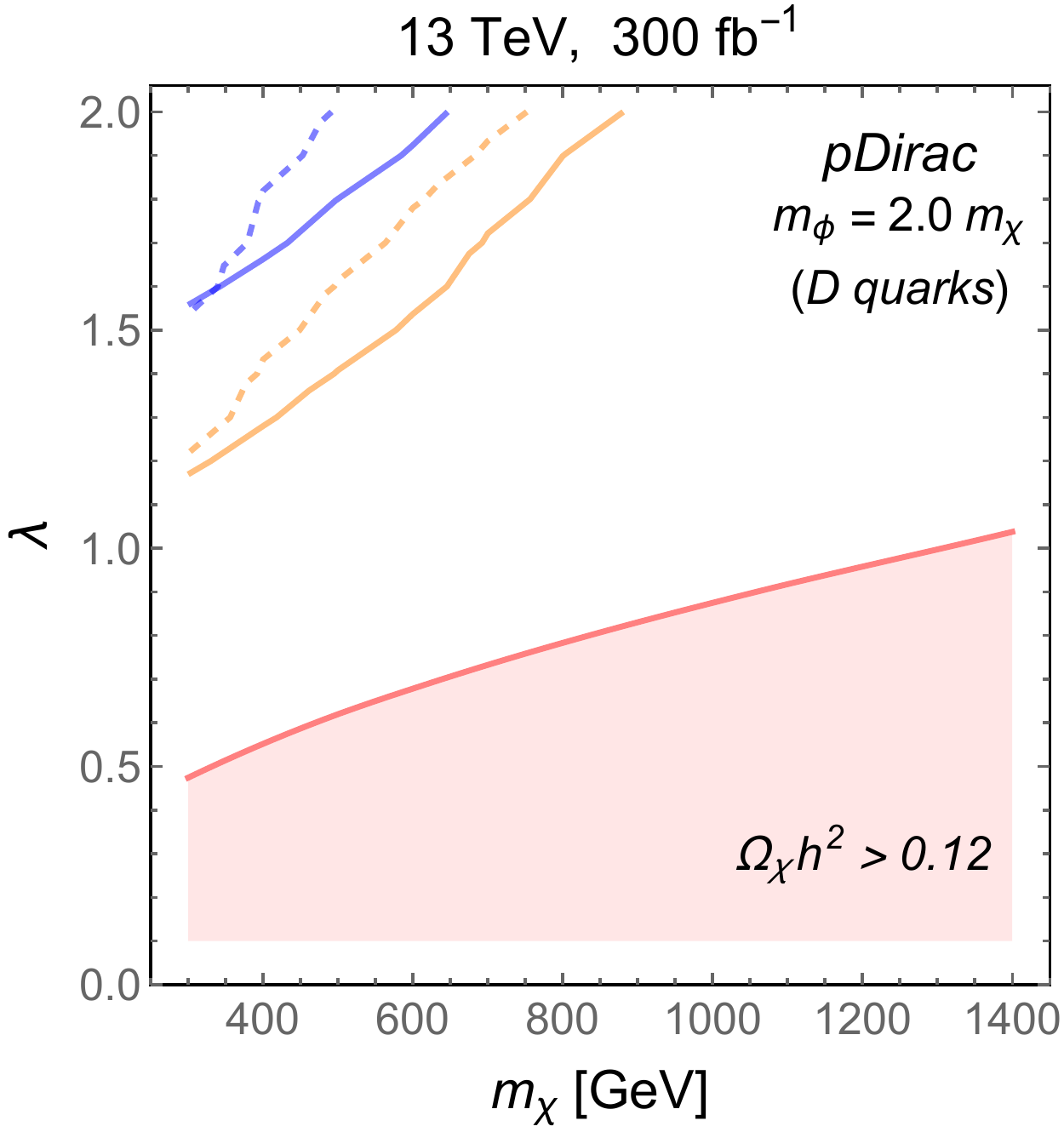}
\end{center}
\caption{
The 95\% \CL reach of our models at the 13 TeV \LHC with 100 fb$^{-1}$.
The color code is as in Fig.~\ref{fig:bounds}.
The dashed blue curves would exclude regions occupied by the blue dots.
}
\label{fig:projections}
\end{figure*}

\begin{figure*}[t!]
\begin{center}
\includegraphics[width=0.45\textwidth]{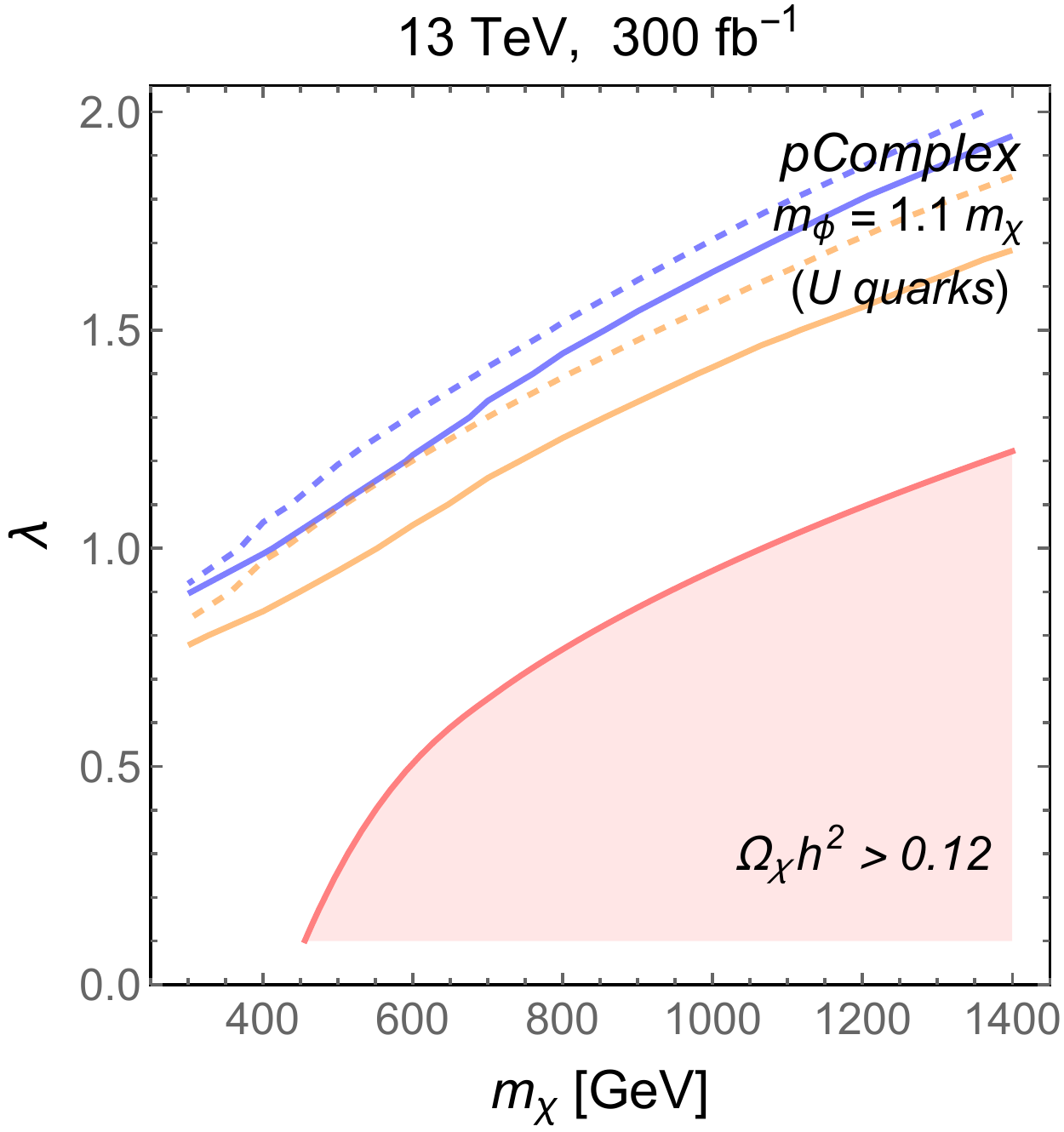}
\quad \quad
\includegraphics[width=0.45\textwidth]{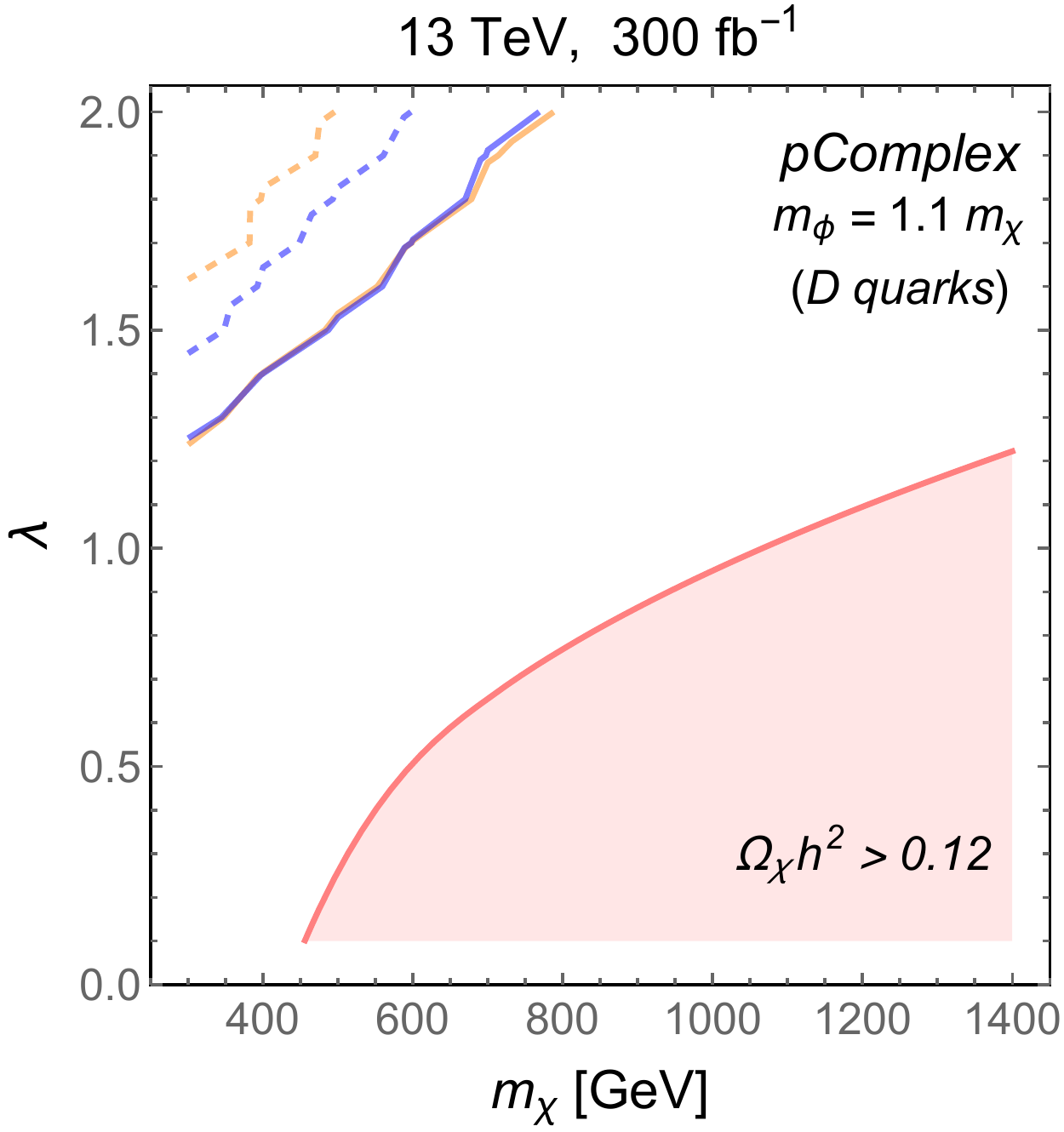}
\\ 
~\\
\includegraphics[width=0.45\textwidth]{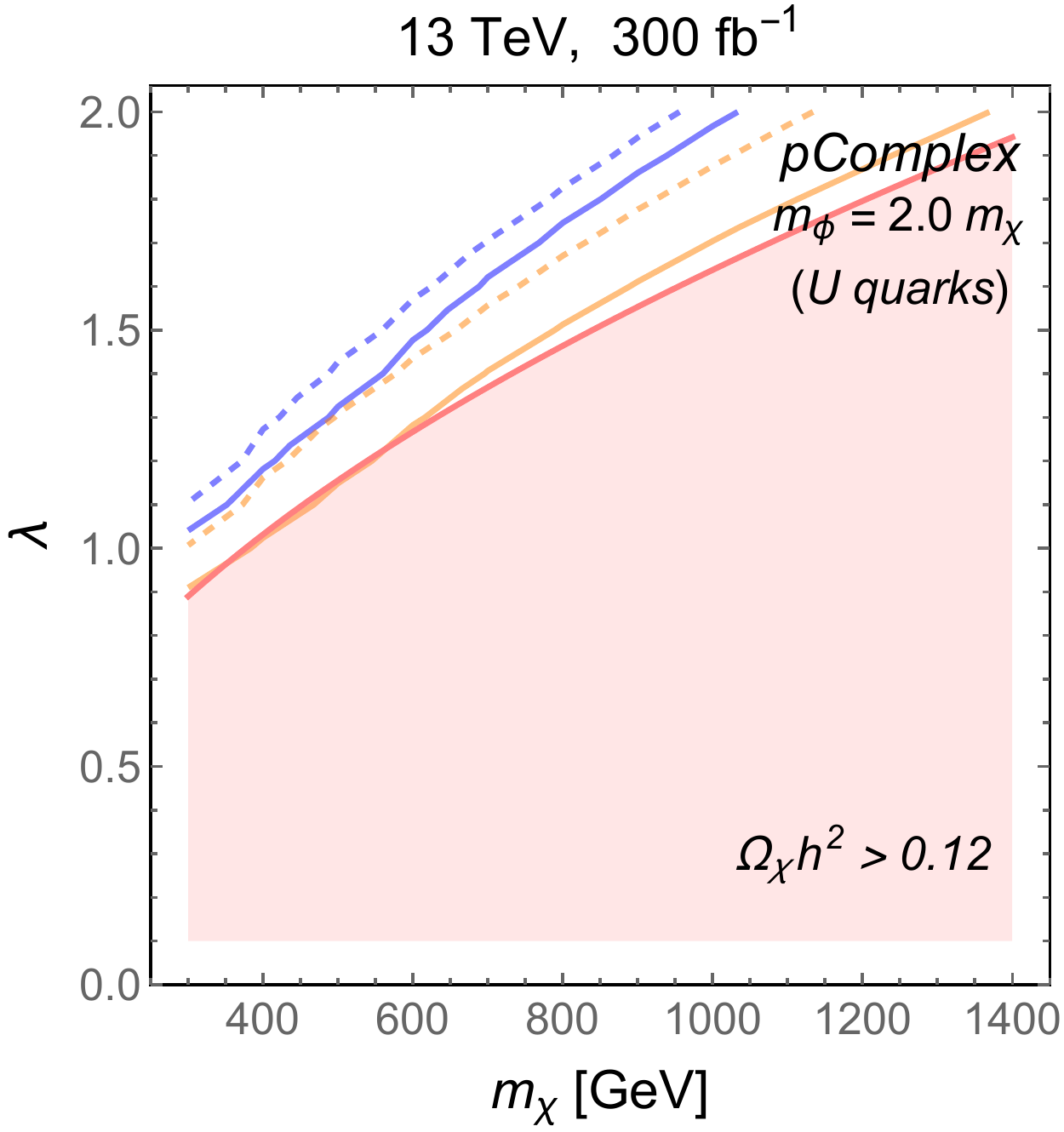}
\quad \quad
\includegraphics[width=0.45\textwidth]{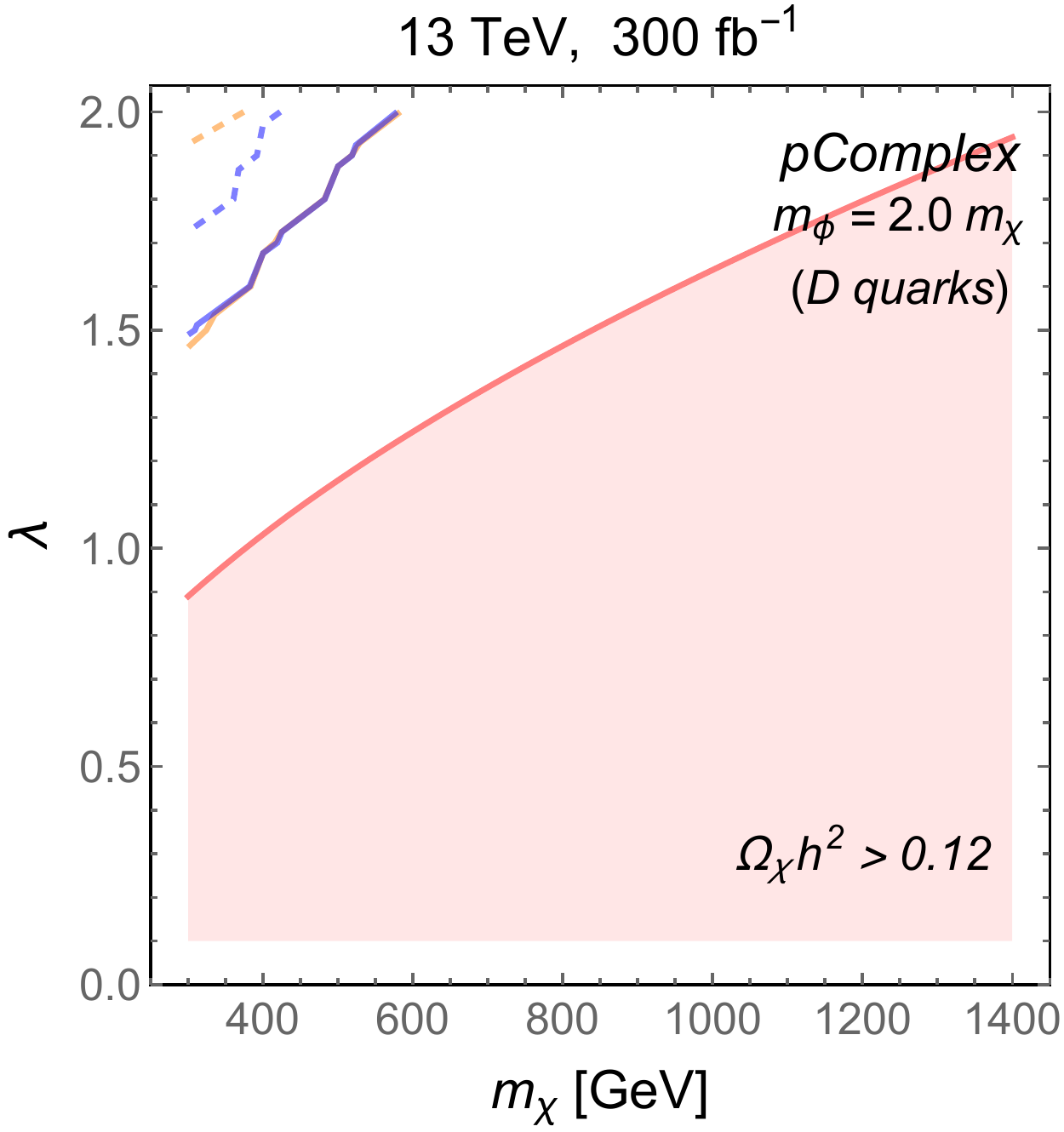}
\end{center}
\caption{
The 95\% \CL reach of our models at the 13 TeV \LHC with 100 fb$^{-1}$.
The color code is as in Fig.~\ref{fig:bounds}.
}
\label{fig:projectionsdown}
\end{figure*}

Our dileptonic probes are highly complementary to jets + $\met$ searches.
At $\mphi = 1.1 \mdm$ the latter are generally weak, since in this compressed region only a small fraction of events pass the tight cuts applied on missing energy.
Consequently, the dileptonic limits for spin-1/2 \DM are seen to generally surpass the jets + $\met$ limits.
This is true of $\sRR$ as well, except the bound on the coupling now rapidly tightens at $\mdm \simeq$ 330 GeV.
This happens because the production rate of the fermionic mediator (in ${\tt \sRR}$,~${\tt \sRL}$) is higher than the scalar mediator (in ${\tt \fRR}$,~${\tt \fRL}$), and we are able to saturate the \CMS exclusion cross section with pure \QCD production ($\lambda \ra 0$) in this region.
As for $\sRRd$, the dilepton bounds are weaker due to small down quark PDFs.
At $\mphi = 2\mdm$ the jets + $\met$ limits are comparable to those at $\mphi = 1.1\mdm$.
This is because, though the signal $\met$ acceptance improves in the uncompressed region, the mediator production rates fall with $\mphi$.
As the dilepton signal is diminished in this region, it complements jets + $\met$ in a model-dependent fashion: for spin-1/2 \DM, the $\mee$ ($\ctCS$) bound outdoes jets + $\met$ for $\fRR$ ($\fRRd$), while for spin-0 \DM, no dilepton bound surpasses jets + $\met$.

The relic density constraint on \DM overproduction is generally stronger for spin-0 \DM than spin-1/2 \DM.
This is because the $s$-wave piece of complex scalar \acro{DM} annihilation is chirality suppressed, whereas that of Dirac \DM is not \cite{1305.1611,1307.8120}.
At $\mmed = 1.1~\mdm$ and $\mdm \leq 450$~GeV, spin-0 \DM gives weaker bounds since the efficient self-annihilation of the colored fermion mediator drives the co-annihilation mechanism in this region.

Our dilepton probes greatly complement direct detection searches as well.
Limits from the latter are generally strong when the mediator is near-degenerate with \DM in mass, as seen in the $\mmed = 1.1\, \mdm$ plots.
This is due to the factor of $(\mmed^2 - \mdm^2)^{-k}$ in the cross sections, where $k$ = 4 (2) for spin-1/2 (spin-0) \DM.
The limit on spin-1/2 \DM is mostly insensitive to $\lambda$ at $\mdm \simeq 600~$GeV due to our scaling of the scattering cross sections with $\Omega_\chi h^2/0.12 \propto \sigmaveeave$ -- the former $\propto \lambda^4$ and the latter $\propto \lambda^{-4}$ at large $\lambda$, where co-annihilations with the mediators are unimportant.
The limit does vary with the coupling at small $\lambda$, where co-annihilations dominate.
This asymptotic behavior of the limit with respect to $\lambda$ allows our dilepton probes to constrain our set-up better than direct detection at $\mdm \gsim 600$~GeV.
On the other hand, the direct detection limit does not asymptote as quickly for spin-0 \DM.
This is because, as just mentioned, \DM annihilations are chirality-suppressed in the $s$-wave, allowing co-annihilations to influence freeze-out even at large $\lambda$. 
As a result, direct detection limits dwarf all other constraints for spin-0 \DM at $\mphi = 1.1 \mdm$.      
The potency of dilepton probes is better at higher $\mmed$.
Due to the $\mphi^{-k}$ scaling, the limits weaken with $\mmed$ so much as to disappear from the $\mmed = 2 \mdm$ plots, allowing dilepons to probe this region better.

Finally, in Figs~\ref{fig:projections} and \ref{fig:projectionsdown} we show the future 95\% \CL sensitivity of our dilepton probes at the 13 TeV \LHC with a luminosity of $\Lc = 300\ {\rm fb}^{-1}$.
The color code is as in Fig.~\ref{fig:bounds}.
To obtain these sensitivities, we performed a $\chi^2$ fit to the background, i.e. we set $N_{d_i} \ra N_{b_i}$ in Eq.~(\ref{eq:chisq}), assuming a systematic error of 2\% and reusing our 8 TeV lepton reconstruction efficiencies.
Our sensitivities improve with respect to the 8 TeV measurements with 20 fb$^{-1}$ luminosity.

This results not only from the usual effect of obtaining better statistics from increase in luminosity, but also crucially, from the increase in {\em collider energy} as well. 
The latter effect originates in the PDFs of the $q\bar{q}$ initial state; at a given $\mee$, their parton luminosity increases with an increase in $\sqrt{s}$.
From this follows the otherwise surprising result that, at the 13 TeV \LHC, $\mee$ measurements surpass $\ctCS$ in sensitivity for all our models.
By populating the $\mee$ spectrum with more events, the increase in parton luminosities magnifies the interference signals seen in Fig.~\ref{fig:signalmll} and contributes more to the $\chi^2$ of the $\mee$ spectrum, while 
the same effects need not be apparent in the $\ctCS$ spectrum, which integrates over a wide range of invariant masses, 400 GeV $\leq \mee \leq$ 4.5 TeV\footnote{Better sensitivities to the interference effects may be obtained by optimizing this $\mee$ window.
Another way to look for these effects, which we do not pursue here, is to use the $\mee$-dependent $\afb$ and $\ace$ (defined in Eqs.~\ref{eq:afb} and \ref{eq:ace}).
}.

In the left-hand panel of Fig.~\ref{fig:projections} the blue dots populate the exclusion regions of the dashed blue curves.
These ``islands" of exclusion form because the model $\fRL$ allows for destructive interference between the \DM box and \SM tree amplitudes, which affects the $\ctCS$ spectrum in peculiar ways.
At large $\lambda$ the signal spectrum is higher than the background, but as we dial $\lambda$ down the signal spectrum approaches the background and eventually crosses it, giving a deficit in events.
The same behavior is seen as we dial the \DM mass up.
As a result, the $\chi^2$ bound first gets weaker and then stronger again as we scan from top to bottom in $\lambda$ or left to right in $\mchi$.

An interesting prospect emerges for $\sRR$ at 400 GeV $\leq \mdm \leq$ 600 GeV.
It can be seen from the bottom left panel of Fig.~\ref{fig:projectionsdown} that in this mass range the $\mll$ sensitivity roughly coincides with the thermal line.
This implies that, should this scenario be realized in nature, the \LHC is poised to find all of the cosmological \DM in the Drell-Yan process.

\section{Conclusions}
\label{sec:disc}

In this work we have shown that dark matter may be characterized using invariant mass and scattering angle spectra of dilepton distributions at the \LHC.
If \DM coupled to both quarks and leptons through $t$-channel mediators, radiative corrections from this dark sector, in combination with threshold effects, produce unique spectral features that may single out \DM properties.
We have shown these features in Figs.~\ref{fig:signalmll}, \ref{fig:signalcostheta} and \ref{fig:signalafbace}.
Our findings can be summarized thus:
\begin{enumerate}
\item Finding a dileptonic signal as sketched in Fig.~\ref{fig:signalmll} would imply that \DM is {\em not} its own anti-particle.
\item The spin of \DM is determinable from the $\mll$ spectrum.
In the region where the signal cross section rises quickly, its slope must be inspected; at higher $\mll$, one must check whether the event ratio of signal over background grows rapidly or settles to a steady value.
Signals would be visible in the angular spectrum, but untangling the spin is more challenging. 
\item If \DM is a fermion, the signal cross section rises abruptly near twice the \DM mass, thus revealing the mass of \DM.
This feature must be reflected as an abrupt change of slope at the same $\mll$ in angular asymmetries, such as the forward-backward ($\afb$) or center-edge asymmetry ($\ace$).
\item The angular spectrum pinpoints the chirality of the fermions to which \DM couples, an effect best seen in the angular asymmetries $\afb$ and $\ace$.
The $\mll$ spectrum picks out this difference poorly.
\end{enumerate}

Having analyzed the signal features, we then placed constraints on the couplings and masses that we introduced using \LHC dilepton data, and contrasted them against bounds from multijets + $\met$ searches, relic density measurements, and direct detection.
We found that angular distributions sometimes gave better constraints than $\mll$ distributions, significantly updating the conclusions of Ref.~\cite{1411.6743}.
We also found that dileptonic measurements in general are complementary to conventional \DM searches.
This is especially true of \DM coupling to up quarks, where these probes often set the strongest collider bound on the {\tt RR} model.
It must be remembered that, just like jets + $\met$, the Drell-Yan process sets bounds not only on \DM but on any analogous neutral particle that lives longer than collider time scales. 

The dilepton sensitivities to our models would increase in future \LHC runs.
The 13 TeV \LHC with a luminosity of $300~{\rm fb}^{-1}$ is poised to cover couplings down to $\lambda \simeq$ 1 for $\mdm \geq 300$~GeV, and \DM masses up to $\mdm$ = 1400 GeV for $\lambda \leq 2$.
A potential hint with sufficient signal significance in these regions would enable us to infer some subset of the \DM properties discussed above.
This could be valuable to complementary experiments such as direct detection and \MET-based collider searches.

For the sake of illustration, we only considered \DM couplings to right-handed quarks.
Should one of our dilepton signals arise in forthcoming runs of the \LHC, we must also entertain interpretations of \DM coupling to other flavors and chiralities of quarks, and consider a wider range of splittings $\delm$ when shape-fitting.
Disentangling the exact Lagrangian structure would be a challenging task, and may involve deeper scrutiny of all available spectral information.

All in all, we look forward to the amusing prospect of visible particle production educating us on dark matter.

\section*{Acknowledgments}

We thank Spencer Chang for a significant conversation. 
We also thank 
Wolfgang Altmannshofer,
Joe Bramante,
Sally Dawson,
Paddy Fox, 
Roni Harnik, 
Mike Hildreth,
Jeff Hutchinson,
and
Graham Kribs for fruitful discussions.
\acro{R.C.} thanks Yuval Grossman and Chris Kolda for illuminating discussions on a key result, 
and 
Asher Berlin,
Thomas Hahn,
Gopolang Mohlabeng
and
Hiren Patel 
for help with some {\tt Mathematica} packages.
This work was partially supported by the National
Science Foundation under Grants 
No. PHY-1417118 and No. PHY-1520966.

\appendix

\begin{figure*}[t!]
\begin{center}
\includegraphics[width=0.45\textwidth]{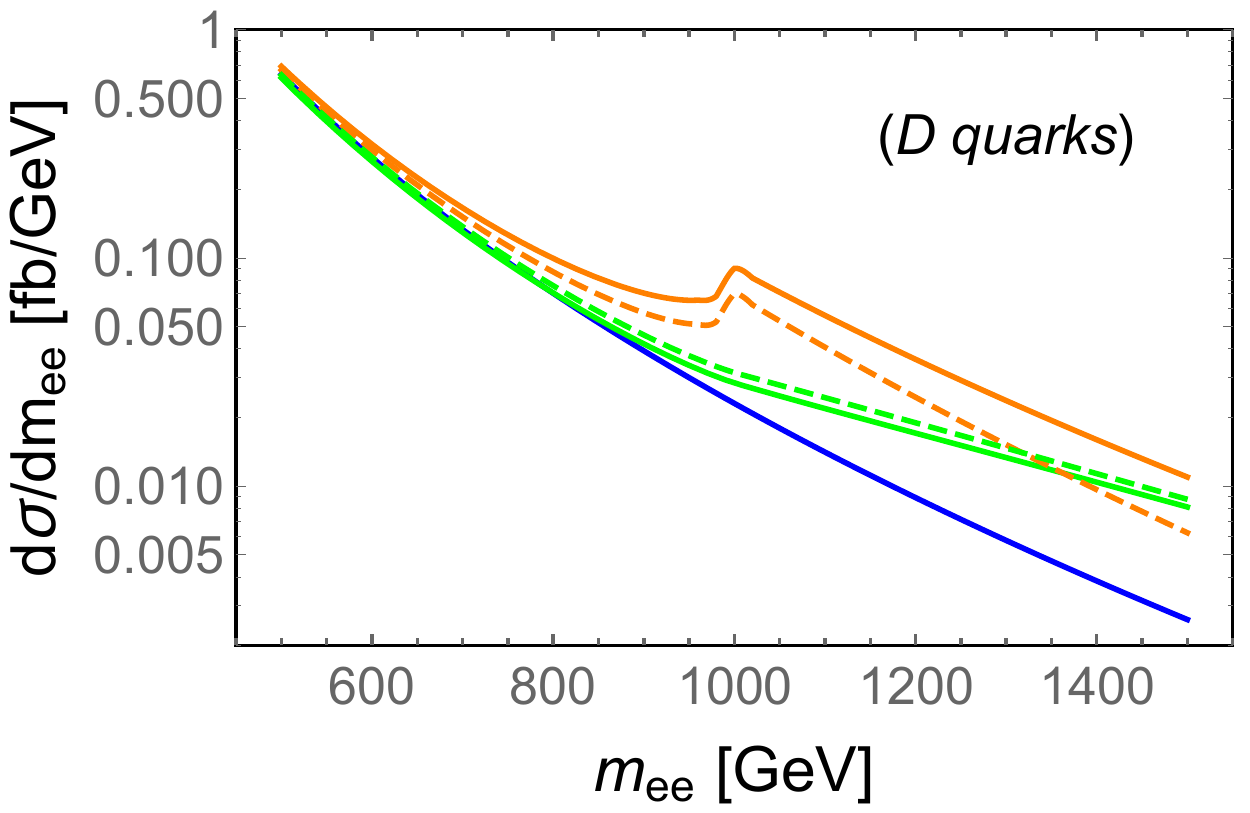}
\quad \quad
\includegraphics[width=0.45\textwidth]{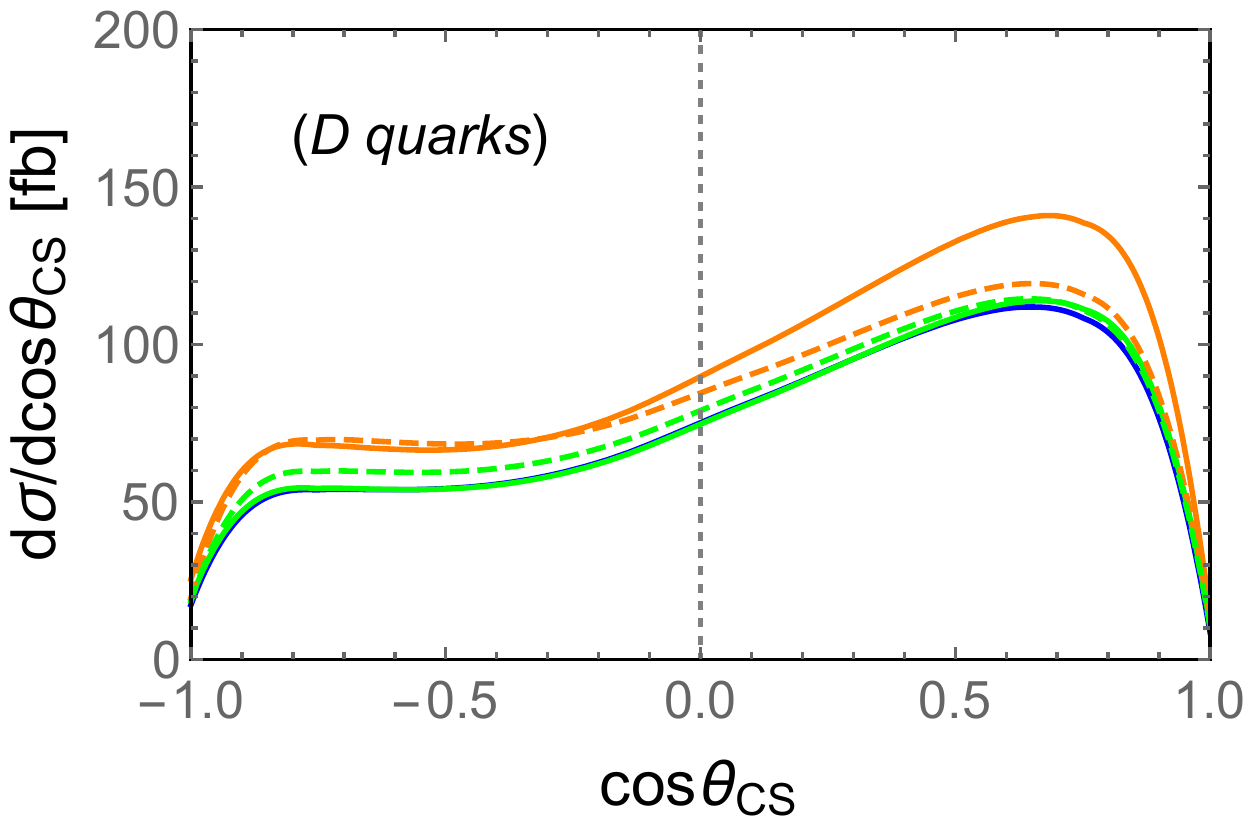}
\\ 
~\\
\includegraphics[width=0.45\textwidth]{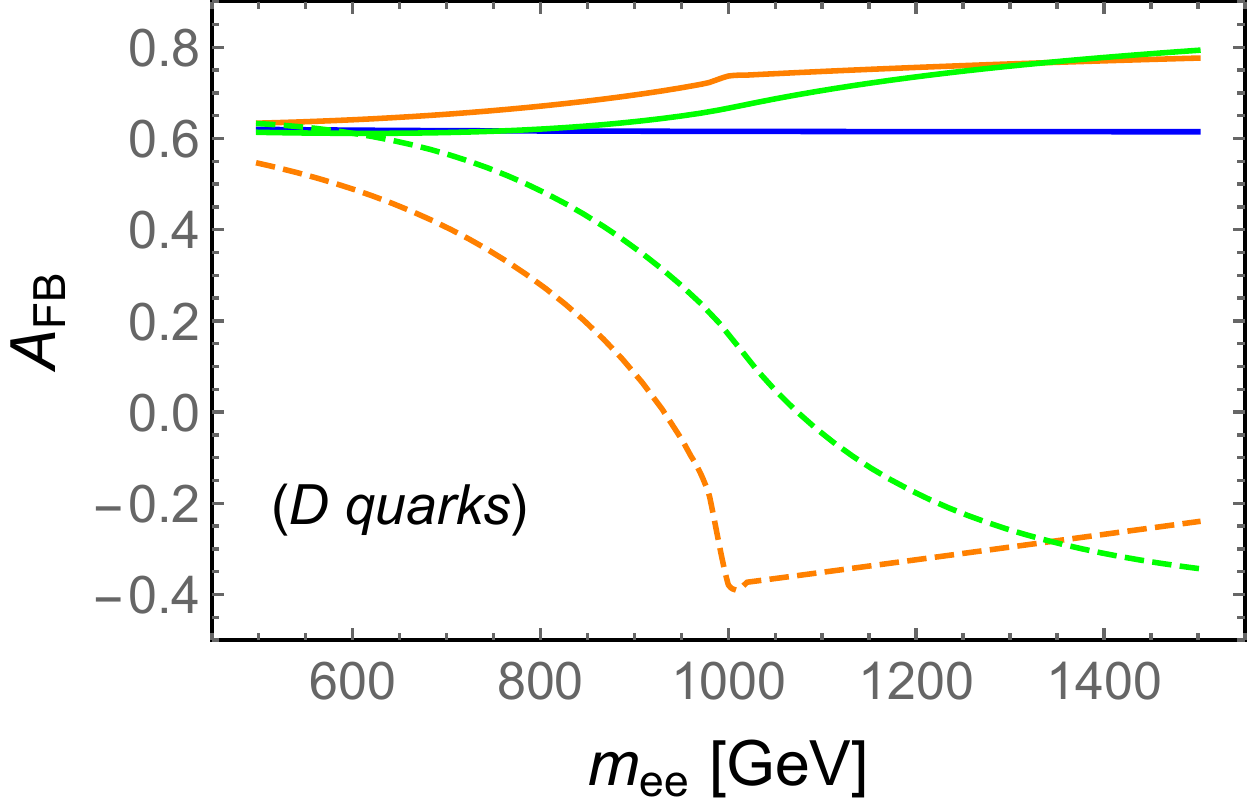}
\quad \quad
\includegraphics[width=0.45\textwidth]{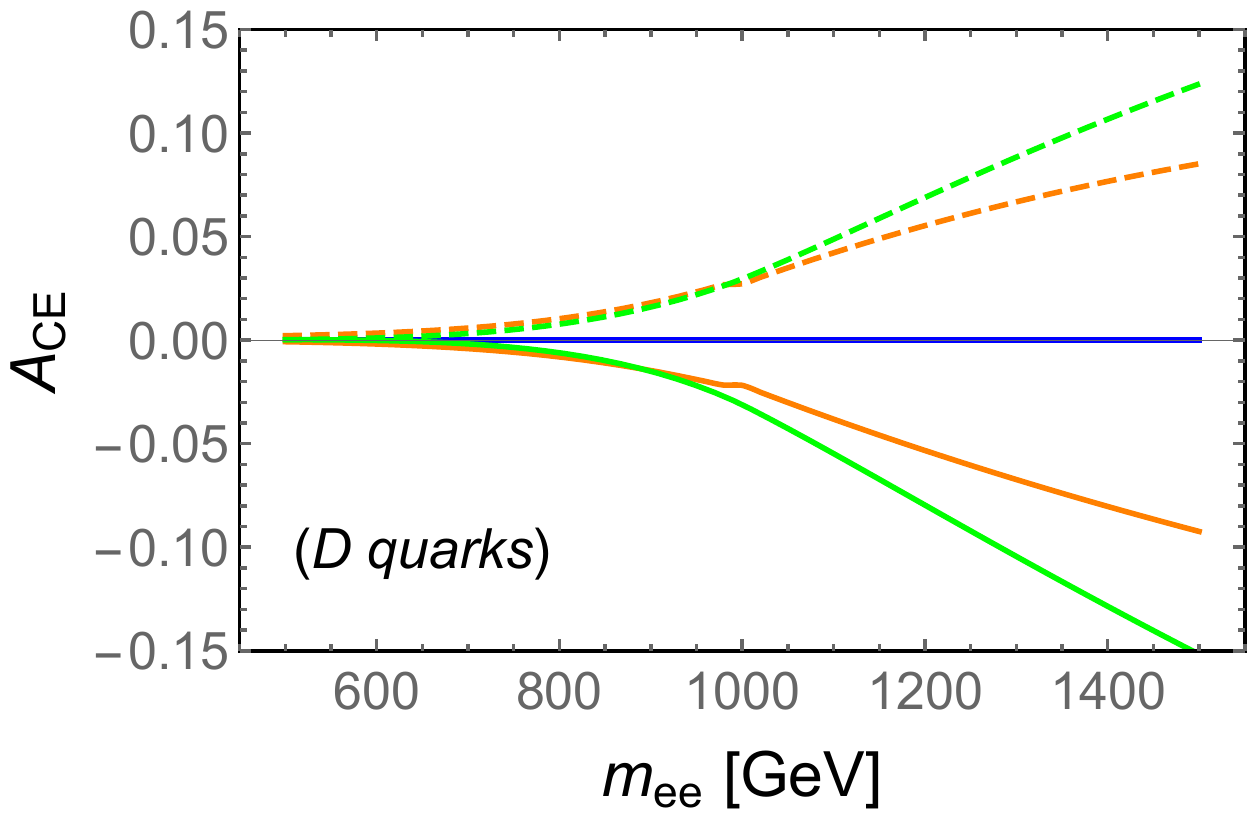}
\end{center}
\caption{
Dilepton signals for the models in which \DM couples to down quarks.
The benchmark point chosen and the color code of these plots is the same as in Sec.~\ref{sec:signals}.
See Appendix~\ref{app:downbenchmark} for more details.}
\label{fig:benchmarksdown}
\end{figure*}

\section{Down quark benchmark plots}
\label{app:downbenchmark}

In this appendix we briefly discuss our dilepton signals for \DM coupling to down quarks.
These signals, in the $\mll$ and $\ctCS$ spectra as well as in the forward-backward and center-edge asymmetries, are plotted in Fig.~\ref{fig:benchmarksdown} with the color code and benchmark point used in Sec.~\ref{sec:signals}. 

The signal features are qualitatively the same as those for \DM coupling to up quarks, hence the discussion in Sec.~\ref{sec:signals} about disentangling \DM properties using these signals holds in this scenario as well.
Whatever visible differences there are arise from the sign of the interference term $d\sigma_{\rm int}$ in Eq.~\ref{eq:XSdefs}.
For the down quark models, this term is always positive, resulting in $d\sigma_{\rm tot} > d\sigma_{\rm SM}$ for all $\mll$.
Thus, while the $\mll$ spectra in Fig.~\ref{fig:signalmll} for the models $\fRR$ and $\fRL$ show a slight deficit with respect to the \SM for $\mll < 2 \mdm$, the analogous models $\fRRd$ and $\fRLd$ produce no such deficits in Fig.~\ref{fig:benchmarksdown}.
Similarly, while the deficits led to a subdued $\ctCS$ signal in Fig.~\ref{fig:signalcostheta}, the analogous signals in Fig.~\ref{fig:benchmarksdown} rise visibly above the background.
The deficits had also led to the signal $\afb$ crossing the \SM $\afb$ in Fig.~\ref{fig:signalafbace}, but in Fig.~\ref{fig:benchmarksdown} the chiralities of the lepton are more neatly divided.
Finally, no interesting difference exists between the $\ace$ in Figs~\ref{fig:signalafbace} and \ref{fig:benchmarksdown}.

\section{Cross section formulae}
\label{app:formulae}

This appendix provides formulae for the dilepton production cross sections $d\sigma_{\rm int}$ and $d\sigma_{\rm \chi}$ in Eq.~(\ref{eq:XSdefs}), and for the \DM-nucleon scattering cross section in direct detection.

\subsection{Dilepton production}

It is convenient to define the following short-hand notation for the Passarino-Veltman (\acro{PV}) box functions:
\beq
D_i \equiv D_i [m_{q}^{2},m_{q}^{2},m_{\ell}^{2},m_{\ell}^{2},\hat{s},\hat{t},\mu_{1}^{2},\mphi^{2},\mu_{2}^{2},\mphi^{2}],
\label{eq:Dij}
\eeq
where $i$ is the \acro{PV} index, and $\mu_{1,2}$ are the \DM eigenmasses in Eqs.~\ref{eq:massfermion} and \ref{eq:massscalar}.
Since in Sec.~\ref{sec:pheno} we had set $\mu_2 - \mu_1$ = 1 MeV, which is unresolvable at the \LHC, we may well approximate $\mu_1 = \mu_2 = \mdm$.

In general, the interference between the tree-level and box amplitudes can be split into a piece in which the tree-level diagram has an $s$-channel--mediated photon, and another in which it has an $s$-channel--mediated $Z$ boson: 
$d\sigma_{\rm int} = d\sigma_{\gamma-{\rm box}} + d\sigma_{Z-{\rm box}}$.
And as explained in Sec.~\ref{sec:signals}, we can approximate $d\sigma_{\rm \chi}$ with the cross-section coming from the squared box amplitude, $d\sigma_{\rm box-box}$.
In the following we provide expressions for these cross sections for our various models, up to a proportionality factor $(32 \pi \mll^2 N_c)^{-1}$.
Here $e$ is the \acro{QED} coupling, $Q_q=2/3 (-1/3)$ is the electric charge of up-type (down-type) quarks, $g$ and $c_W$ are the electroweak coupling and mixing angle respectively, $a_f$ and $b_f$ are respectively the vectorial and axial couplings between SM fermions and the $Z$ boson, and $m_Z$ and $\Gamma_Z$ are the mass and the width of the $Z$ boson.
These expressions were obtained using {\tt FeynCalc} \cite{Mertig:1990an,1601.01167}, and numerical results obtained with {\tt LoopTools} \cite{9807565} and {\tt Package-X} \cite{1612.00009}.

\subsubsection{$\fRR, \fRRd, \sRR, \sRRd$}
\bea
\nn d\sigma_{{\rm \gamma-box}} &\propto& -\frac{e^{2}Q_{q}\left(\hat{s}+\hat{t}\right)^{2}\lambda^{4}}{4\pi^{2}\hat{s}}{\rm Re}[\widetilde{\mathcal{D}}], \\
\nn d\sigma_{Z-{\rm box}} &\propto& -\frac{g^{2}(a_{\ell}-b_{\ell})(a_{q}-b_{q})\left(\hat{s}+\hat{t}\right)^{2}\lambda^{4}}{16\pi^{2}c_{W}^{2}\left[\left(m_{Z}^{2}-\hat{s}\right)^{2}+m_{Z}^{2}\Gamma_{Z}^{2}\right]} \\
\nn & & \times{\rm Re}\left[\left(m_{Z}^{2}-\hat{s}-im_{Z}\Gamma_{Z}\right)\widetilde{\mathcal{D}}\right], \\
\nn d\sigma_{{\rm box-box}} &\propto& \frac{\left(\hat{s}+\hat{t}\right)^{2}\lambda^{8}}{64\pi^{4}}|\widetilde{\mathcal{D}}|^2,
\label{eq:box2 RR}
\eea

The $\widetilde{\mathcal{D}}$ are combinations of \acro{PV} functions: 
\bea
\nn \widetilde{\mathcal{D}}_{\tt pD^{u,d}_{\rm RR}} &=& 2D_{00}+\hat{s}\left(D_{2}+D_{12}+D_{22}+D_{23}\right), \\
\nn \widetilde{\mathcal{D}}_{\tt pCS^{u,d}_{\rm RR}} &=& 2D_{00}-\hat{t}D_{13}~.
\eea
\subsubsection{$\fRL, \fRLd, \sRL, \sRLd$}
\bea
d\sigma_{{\rm \gamma-box}} &\propto& -\frac{e^{2}Q_{q}\hat{t}^{2}\lambda^{4}}{4\pi^{2}\hat{s}}{\rm Re}[\widetilde{\mathcal{D}}], \\
\nn d\sigma_{Z-{\rm box}} &\propto& -\frac{g^{2}(a_{\ell}+b_{\ell})(a_{q}-b_{q})\hat{t}^{2}\lambda^{4}}{16\pi^{2}c_{W}^{2}\left[\left(m_{Z}^{2}-\hat{s}\right)^{2}+m_{Z}^{2}\Gamma_{Z}^{2}\right]} \\
& & \times{\rm Re}\left[\left(m_{Z}^{2}-\hat{s}-im_{Z}\Gamma_{Z}\right)\widetilde{\mathcal{D}}\right], \\
d\sigma_{{\rm box-box}} &\propto& \frac{\hat{t}^{2}\lambda^{8}}{64\pi^{4}}|\widetilde{\mathcal{D}}|^2~,
\eea
with
\bea
\nn \widetilde{\mathcal{D}}_{\tt pD^{u,d}_{\rm RL}} &=& \mchi^{2}D_{0}~, \\
\nn \widetilde{\mathcal{D}}_{\tt pCS^{u,d}_{\rm RL}} &=& 2D_{00}-\left(\hat{s}+\hat{t}\right)D_{13}~.
\eea

\subsection{Direct detection}
At direct detection experiments, our \DM behaves like a Majorana or real scalar particle (see Sec.~\ref{sec:models}).
Here we provide the appropriate spin-independent per-nucleon scattering cross sections, $\sigma_{\rm SI}$.

\subsubsection{Majorana DM ($\fRR, \fRL, \fRRd, \fRLd$)}
We have
\bea
\nn \sigma_{\rm SI}=\frac {4}{\pi}\mu_{\chi N}^2|f_N|^2~,
\eea
where $\mu_{\chi N}$ is the \DM-nucleon ($N=p,n$) reduced mass, and effective coupling $f_N$ is given by \cite{1307.8120}
\bea
\nn \frac {f_N}{m_N}= f_q f_{T_u} + \frac{3}{4}(q_2+\bar{q}_2)g_q - \frac{8\pi}{9\alpha_s}f_{T_G}f_G~,
\eea
with the Wilson coefficients
\bea
f_q &=& \frac{\mchi\lambda^2}{8(\mphi^2-\mchi^2)^2}, \\
f_G &=& -\frac{\alpha_s\mchi\lambda^2}{96\pi \mphi^2(\mchi^2-\mchi^2)^2},
\eea
and the coefficients
$f_{T_u}$(proton) = 0.023,
$f_{T_u}$(neutron) = 0.017,
$f_{T_d}$(proton) = 0.032,
$f_{T_d}$(neutron) = 0.041,
$u_2=0.22$,
$\bar{u}_2 = 0.034$,
$d_2=0.11$,
$\bar{d}_2 = 0.036$,
$g_q=4 f_q$,
$f_{T_G}$(proton) = 0.925,
$f_{T_G}$(neutron) = 0.922 \cite{1012.5455,1312.4951}.

\subsubsection{Real Scalar DM ($\sRR, \sRL, \sRRd, \sRLd$)}

Here
\bea
\nn \sigma_{\rm SI} = \frac {\mu_N^2}{\pi}\left(\frac{f_N}{\mchi}\right)^2~,
\eea
with
\bea
\frac {f_N}{m_N}= f_q f_{T_q} + \frac{3}{4}(q_2+\bar{q}_2)g_q.
\eea
The coefficients on the right-hand side are the same as before.

\section{Other constraints}
\label{app:otherconstraints}

Here we compile experiments that are relevant to our models, but place constraints so weak as to not appear in our plots in Sec.~\ref{sec:pheno}.
\\
\subsection{LEP}
Box diagrams similar to those in Fig.~\ref{fig:Feyn} but involving only $\lmed$ will generate four-lepton contact operators, which will contribute to $e^+ e^- \rightarrow \ell^+ \ell^-$.
\acro{LEP} measurements of this process can thus constrain our parameters.
It was found in \cite{1411.6743} that for the spin-1/2 \DM models $\fRR$ and $\fRRd$, the \acro{LEP} limit was $\mdm \gsim 250~$GeV for $\lambda \lsim 2$, in agreement with \cite{1408.1959}.
A similar limit holds for $\fRL$ and $\fRLd$.
And as shown in \cite{1408.1959}, the limits are even weaker for spin-0 \DM ($\sRR$, $\sRRd$, $\sRL$ and $\sRLd$), with $\mdm \gsim 200~$GeV for $\lambda \lsim 2$. 
\subsection{Muon anomalous magnetic moment}

Due to the presence of leptonic mediators, our models would contribute to $(g-2)_\mu$ through loops if \DM interacted with the muon.
Using formulae from \cite{1402.7358}, we find that our spin-1/2 \DM models widen the long-standing 3$\sigma$ discrepancy between the \SM prediction and the measurement.
If we require $< 5\sigma$ deviation from the measurement, the limit is $\mdm \simeq \mphi \gsim 200$~GeV for $\lambda = 2$, which is a much weaker limit than Xenon1T.
On the other hand, our spin-0 \DM models contribute in the direction of the measurement; the discrepancy is explained at $\mdm \simeq \mphi = 175~$GeV for $\lambda = 2$ and at smaller masses for smaller couplings.
However, these values are already ruled out by Xenon1T.  

\subsection{Fermi-LAT}

Since all our annihilations are $s$-wave, our scenarios can be potentially probed at indirect detection searches looking for present-day annihilation of \DM.
The most stringent constraints are set by Fermi-\acro{LAT} observations of dwarf galaxies \cite{1310.0828} that look for \DM annihilation-induced $\gamma$-ray flux.
This flux $\propto \rho_{\rm \DM}^2 \sigmaveeann$.
Thus, if our model populates a fraction $f \equiv \Omega_\chi h^2/0.12$ of \DM at freezeout, it contributes to a fraction $\simeq f$ of the flux.
Hence we must expect the Fermi-\acro{LAT} limit on our model to be weaker than the limit quoted at the thermal cross section $3 \times 10^{-26} {\rm cm}^3/{\rm s}^{-1}$.  
From \cite{1310.0828}, the latter limit is already $\mdm \gsim 100~$GeV, so we expect our Fermi-\acro{LAT} constraint to be much weaker than the ones discussed in Sec.~\ref{sec:pheno}.

\end{document}